\documentclass[screen, acmlarge]{acmart}

\AtBeginDocument{%
  }

\setcopyright{rightsretained}
\copyrightyear{2023}
\acmYear{2023}
\acmDOI{10.1145/3591469}

\acmBooktitle{ACM Trans. Recomm. Syst.}
\acmPrice{15.00}
\acmISBN{978-1-4503-XXXX-X/18/06}

\usepackage{graphicx}
\usepackage{textcomp}
\usepackage{xcolor}
\usepackage{xspace}
\usepackage{subcaption}
\usepackage{multirow}
\usepackage{caption}
\usepackage{enumitem}  
\usepackage{algorithm, algorithmicx, algpseudocode}

\newcommand{\ie}{\emph{i.e.},\xspace}
\newcommand{\eg}{\emph{e.g.},\xspace}

\begin{document}

\title{SelfCF: A Simple Framework for Self-supervised Collaborative Filtering}


\author{Xin Zhou}
\email{xin.zhou@ntu.edu.sg}
\affiliation{%
	\institution{Nanyang Technological University}
	\country{Singapore}
}

\author{Aixin Sun}
\affiliation{%
	\institution{Nanyang Technological University}
	\country{Singapore}
}

\author{Yong Liu}
\affiliation{%
	\institution{Nanyang Technological University}
	\country{Singapore}
}

\author{Jie Zhang}
\affiliation{%
	\institution{Nanyang Technological University}
	\country{Singapore}
}

\author{Chunyan Miao}
\affiliation{%
	\institution{Nanyang Technological University}
	\country{Singapore}
}


\begin{abstract}
Collaborative filtering (CF) is widely used to learn informative latent representations of users and items from observed interactions.
Existing CF-based methods commonly adopt negative sampling to discriminate different items. 
That is, observed user-item pairs are treated as positive instances; unobserved pairs are considered as negative instances and are sampled under a defined distribution for training.
Training with negative sampling on large datasets is computationally expensive.
Further, negative items should be carefully sampled under the defined distribution, in order to avoid selecting an observed positive item in the training dataset.
Unavoidably, some negative items sampled from the training dataset could be positive in the test set.
Recently, self-supervised learning (SSL) has emerged as a powerful tool to learn a model without negative samples.
In this paper, we propose a self-supervised collaborative filtering framework (SelfCF), that is specially designed for recommender scenario with implicit feedback.
The proposed SelfCF framework simplifies Siamese networks and can be easily applied to existing deep-learning based CF models, which we refer to as backbone networks.
The main idea of SelfCF is to augment the latent embeddings generated by backbone networks instead of the raw input of user/item ids. 
We propose and study three embedding perturbation techniques that can be applied to different types of backbone networks including both traditional CF models and graph-based models. 
The framework enables learning informative representations of users and items without negative samples, and is agnostic to the encapsulated backbones.
We conduct experimental comparisons on four datasets, one self-supervised framework and eight baselines to show that our framework may achieve even better recommendation accuracy than the encapsulated supervised counterpart with a 2$\times$--4$\times$ faster training speed. 
The results also demonstrate that SelfCF can boost up the accuracy of a self-supervised framework BUIR by 17.79\% on average and shows competitive performance with baselines.
\end{abstract}



\maketitle

\section{Introduction}
\label{sec:introduction}
\begin{figure}[t]
	\centering
	\includegraphics[width=0.68\columnwidth,  trim={10 0 10 0},clip]{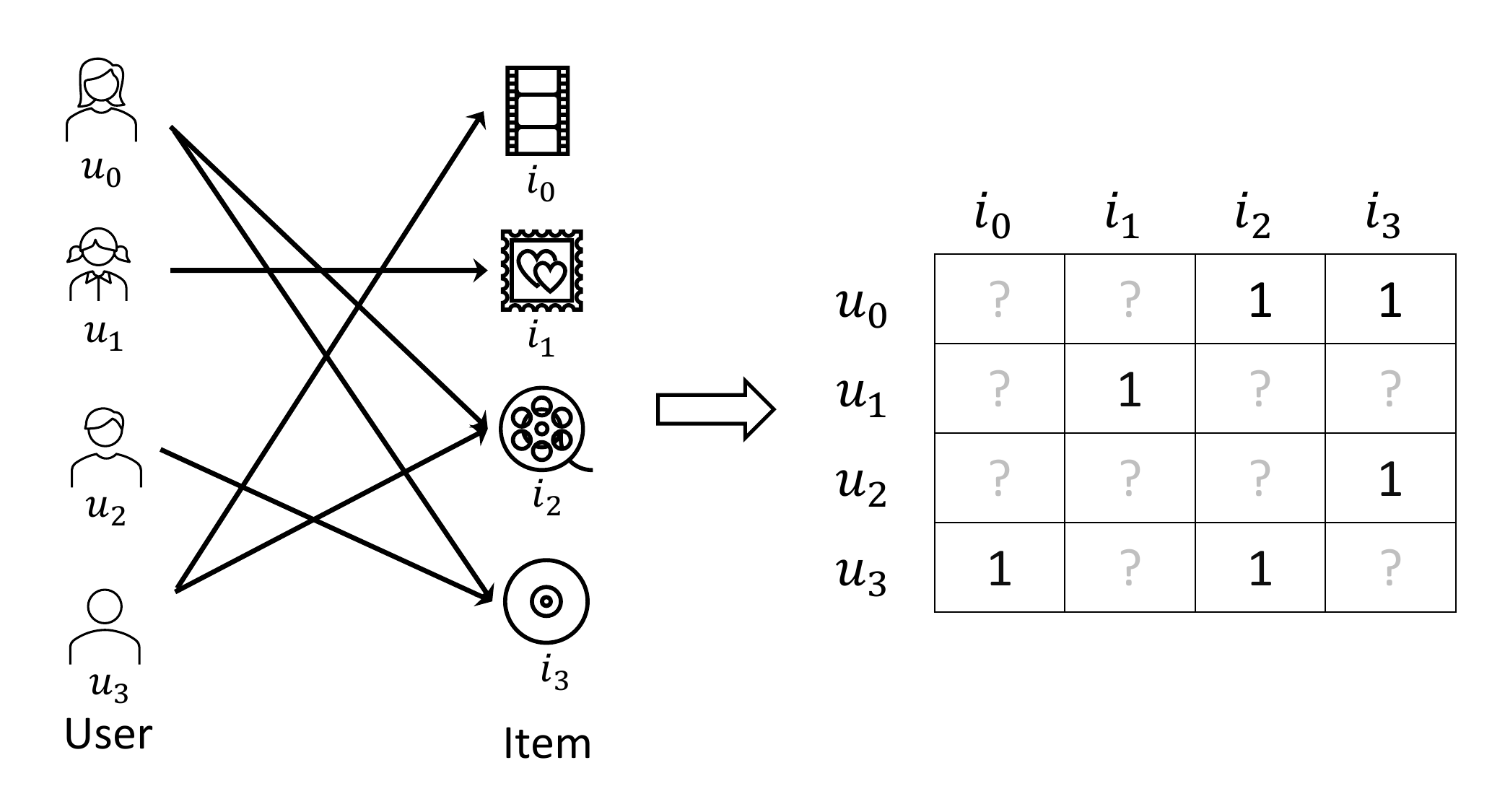}
	\caption{An illustration of the recommendation scenario under implicit feedback. Only positive samples can be captured for training recommender systems.}
	\label{fig:uigraph}
\end{figure}

Recommender systems aim to provide users with personalized products or services. They help to handle the increasing information overload problem and improve customer relationship management. 
In Fig.~\ref{fig:uigraph}, we present an illustration of recommendation under implicit feedback.
Recommender systems are designed to infer the missing values of the matrix (right) transformed from the user-item interactions (left). In top-$K$ scenario, the inferred values are further ranked with each user for personalized recommendation. 
Collaborative Filtering (CF) is a canonical recommendation technique, which predicts interests of a user by aggregating information from similar users or items. 
In detail, existing CF-based methods~\citep{koren2008factorization,shi2014collaborative,he2017neural, he2020lightgcn} learn latent representations of users and items, by first factorizing the observed interaction matrix, then predicting the potential interests of user-item pairs based on the dot-product of learned embeddings.
However, existing CF models rely heavily on negative sampling techniques to discriminate against different items, because negative samples are not naturally available.

Nevertheless, the negative sampling techniques suffer from a few limitations.
Firstly, they introduce additional computation and memory costs.
In existing CF-based methods, the negative sampling algorithm need be carefully designed in order to not select the observed positive user-item pairs. 
Specifically, to sample one negative user-item pair for a specific user, the algorithm checks its conflicts with all the observed positive items interacted with this user.
As a result, much computation is needed for users who have a large number of interactions.
Secondly, even if non-conflicted negative samples are selected for a user, the samples may fall into future positive items of the user. The reason is that the unobserved user-item pairs can be either true negative instances (\textit{i.e.,} the user is not interested in these items) or missing values (\textit{e.g.,} interaction pairs not observed in the training set but the test set)~\cite{rendle2009bpr, lee2021bootstrapping}. We denote the sampled pairs that fall in the test set as false negative samples~\cite{chenrevisiting2022}. 
Although another line of work~\cite{chen2020efficient, chen2020efficienth, chen2020jointly} has get rid of negative sampling and takes the full unobserved interactions as negative samples, they may still treat a future positive sample as negative.

\begin{figure}[bpt]
	\begin{minipage}[t]{0.49\textwidth}
		\centering
		\vspace{0pt}
		\includegraphics[width=0.95\columnwidth,  trim={10 0 10 0},clip]{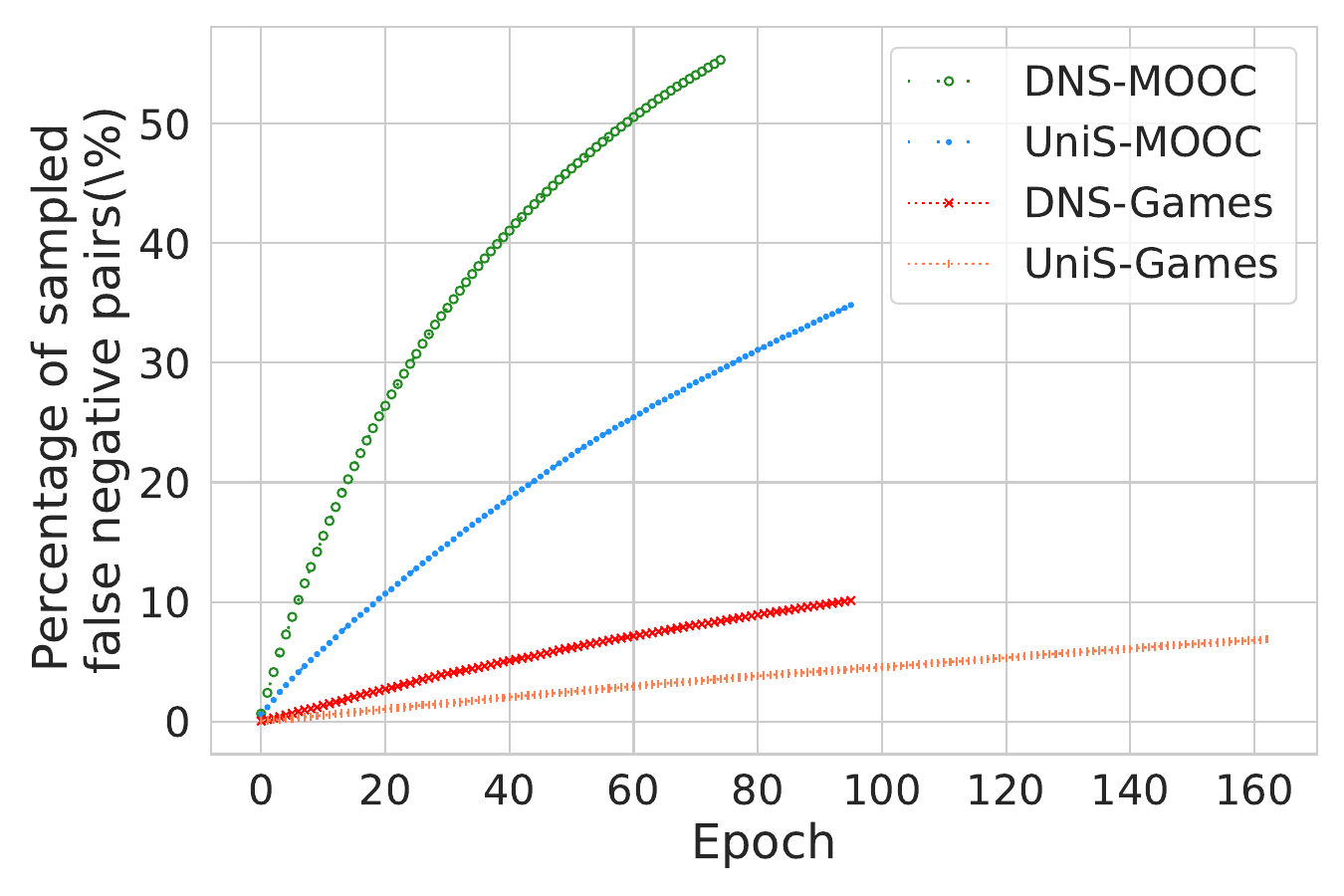}
		\caption{Percentage of false negative pairs sampled with different sampling methods of LightGCN. The percentage is calculated by dividing the number of sampled false negative pairs by the number of instances in the test set and multiplying the result by 100.}
		\label{fig:ns_issues}
	\end{minipage}
	\hfill
	\begin{minipage}[t]{0.49\textwidth}
		\centering
		\vspace{0pt}
		\def\arraystretch{1.15}%
		\setlength\tabcolsep{2.0pt} 
		\captionof{table}{Negative sampling time under various sampling methods and datasets. Overhead is calculated as the percentage of sampling time over training time per epoch. UniS denotes uniform sampling. DNS denotes dynamic negative sampling.}
		\label{tab:ns_issues}
		\begin{tabular}{ccrrr}
			\toprule
			\multirow{2}{*}{Dataset} & \multirow{2}{*}{Method} & \multicolumn{2}{c}{Time ($s$)} & \multirow{2}{*}{Overhead} \\
			& & \multicolumn{1}{c}{Sampling} & \multicolumn{1}{c}{Training} &  \\ 
			\midrule
			MOOC & UniS & 2.32 & 8.89 & 26.1\% \\
			MOOC & DNS & 32.38 & 40.33 & 80.3\%\\
			Games & UniS & 3.47 & 8.22 & 42.2\% \\
			Games & DNS & 39.15 & 43.68 & 89.6\% \\
			\bottomrule
		\end{tabular}
	\end{minipage}
\end{figure}

To uncover the negative sampling problem in current models, we employ uniform sampling (UniS) and Dynamic Negative Sampling (DNS)~\cite{zhang2013optimizing} in LightGCN~\cite{he2020lightgcn} to study the aforementioned issues. 
Uniform sampling is a widely used and classical solution in the item recommendation domain with implicit feedback~\cite{chenrevisiting2022}. DNS improves uniform sampling by selecting a set of negative candidates and ranking the candidates based on learned user/item embeddings. The top-ranked item is used as a hard instance. As a result, DNS is model-sensitive.
We plot the percentage of sampled false negative pairs in the test set along with the training progress of LightGCN under uniform sampling and DNS in Fig.~\ref{fig:ns_issues}.
We test the negative sampling methods on two diverse datasets, MOOC and Amazon Video Games (Games). 
MOOC contains 458,453 interactions collected from 82,535 learners on 1,302 courses. Games has 50,677 users, 16,897 items and 454,529 interactions. 
The sparsity of MOOC and Games are 99.4039\% and 99.9469\%, respectively.
From Fig.~\ref{fig:ns_issues}, we observe the percentage of false negative pairs sampled by DNS is over 50\% when LightGCN early stops on the MOOC dataset. Here, we use the original early stopping setting in LightGCN\cite{he2020lightgcn}. The sparse dataset, Games, has a relatively small number of sampled false negative instances under 10\%. However, the overhead to sample a negative instance increases with the number of candidates, as shown in Table~\ref{tab:ns_issues}.
Although DNS can sample hard negative instances, its overhead on sampling is 2-3 times of uniform sampling in Table~\ref{tab:ns_issues}.
From the above observations, it is promising to train the model without negative sampling.

Self-supervised learning (SSL) models~\cite{grill2020bootstrap, chen2021exploring, zbontar2021barlow} provide us a possible solution to tackle the aforementioned limitations. SSL enables training a model by iteratively updating network parameters without using negative samples. 
Research in various domains ranging from Computer Vision (CV) to Natural Language Processing (NLP), has shown that SSL is possible to achieve competitive or even better results than supervised learning~\cite{grill2020bootstrap, chent2020simple, zbontar2021barlow}. 
The underlying idea is to maximize the similarity of representations obtained from different \textit{distorted versions} of a sample using a variant of Siamese networks~\cite{hadsell2006dimensionality}.
Siamese networks usually include two symmetric networks (\textit{i.e.,} online network and target network) for inputs comparing. The problem with only positive sampling in model training is that, the Siamese networks collapse to a trivial constant solution~\cite{chen2021exploring}. 
Thus, in recent work, BYOL~\cite{grill2020bootstrap} and SimSiam~\cite{chen2021exploring} introduce asymmetry to the network architecture by adding parameter update technique.
Specifically, in the network architecture, an additional ``predictor'' network is stacked onto the online encoder. For parameter update, a special ``stop gradient'' operation is highlighted to prevent solution collapsing.
SimSiam simplifies BYOL by removing its ``momentum update'', which updates the parameters of target networks based on online networks.
We will illustrate the architectures in detail in the related work section.

To the best of our knowledge, BUIR~\cite{lee2021bootstrapping} is the only recommendation framework to learn user and item latent representations without negative samples. 
BUIR is derived from BYOL~\cite{grill2020bootstrap}. Similar to BYOL, BUIR employs two distinct encoder networks (\textit{i.e.,} online and target networks) to address the recurring of trivial constant solutions in SSL. In BUIR, the parameters of the online network are optimized towards that of the target network. At the same time,  parameters of the target network are updated based on momentum-based moving average~\cite{tarvainen2017mean, grill2020bootstrap, he2020momentum} to slowly approximate the online network~\cite{lee2021bootstrapping}.
As BUIR is built upon BYOL, which stems from vision domain, its architecture is redundant and suffers from slow convergence, because of the design of the momentum-based parameter updating.
The SimSiam network is originally proposed in vision domain as well. The input is an image, and techniques for data augmentation on images are relatively mature~\cite{shorten2019survey}, such as random cropping, resizing, horizontal flipping, color jittering, converting to grayscale, Gaussian blurring, and solarization.
As for a pair of user id and item id that is observed in implicit feedback, there is no standard solution on how to distort it while keep its representation invariant.

The learning paradigm of SSL without negative samples differs slightly from existing paradigms that use negative samples to learn representations. SSL without negative samples intends to learn an encoder with augmentation-invariant representation~\cite{grill2020bootstrap, chen2021exploring}. That is, they minimize the representation distance between two positive samples based on a Siamese network architecture~\cite{bromley1993signature}. 
By using negative samples in SSL, solutions are prevented from collapsing because of their repulsivity. Our proposed framework can achieve competitive performance without harnessing this repulsive force.

In this paper, we propose a \textbf{Self}-supervised \textbf{C}ollaborative \textbf{F}iltering (SelfCF) framework, which performs posterior perturbation on user and item latent embeddings, to obtain a contrastive pair.
On architecture design, our framework uses only one encoder instead of two encoders, which simplifies BYOL and SimSiam. 
Besides, instead of perturbing inputs ahead of encoding, we generate different but invariant contrastive views with posterior embedding perturbations.
An additional benefit of posterior embedding perturbation is that the framework can take the internal implementation of the encapsulated backbones as black-box. Conversely, BUIR adds momentum-based parameter updating to encoders in order to generate different views. 
Our experiments on four real-world datasets validate that the proposed SelfCF framework is able to learn informative representation solely
based on positive user-item pairs. 
In our experiments, we encapsulate two popular CF-based models into the framework, and the results on Top-$K$ item recommendation are competitive or even better than their supervised counterparts.

We summarize our contributions as follows:
\begin{itemize}
	\item We propose a novel framework, SelfCF, that learns  latent representations of users/items solely based on positively observed interactions. The framework uses posterior output perturbation to generate different augmented views of the same user/item embeddings for contrastive learning.
	
	\item We design three output perturbation techniques: historical embedding, embedding dropout, and edge pruning to distort the output of the backbone. 
	The techniques are applicable to all existing CF-based models as long as their outputs are embedding-like.
	
	\item We investigate the underlying mechanisms of the framework by performing ablation study on each component. 
	We find the presentations of user/item can be learnt even without the ``stop gradient'' operator, which shows different behaviors from previous SSL frameworks (\textit{e.g.,} BYOL~\cite{grill2020bootstrap} and SimSiam~\cite{chen2021exploring}).
	
	\item Finally, we conduct experiments on four public datasets by encapsulating two popular backbones. Results show SelfCF is competitive or better than their supervised counterpart and outperforms existing SSL framework by up to 17.79\% on average.
\end{itemize}

\section{Related Work}
In this section, we first review the CF technique, then summarize the current progress of SSL.

\subsection{Collaborative Filtering}
CF is a typical and prevalent technique adopted in modern recommender systems~\cite{ying2018graph}. The core concept is that similar users tend to have similar tastes on items. 
To tackle the data sparsity and scalability of CF, more advanced method, Matrix Factorization (MF), decomposes the original sparse matrix to low-dimensional matrices with latent factors/features and less sparsity. 
To learn informative and compressed latent features, deep learning based models are further proposed for recommendation~\cite{wang2015collaborative, he2017neural, zhang2019deep}.

With the emerge of graph convolutional networks (GCNs), which generalize convolutional neural networks (CNNs) on graph-structured data~\cite{zhou2023inductive, liu2023m2gcn, zhang2023dual}, GCN-based CF is widely researched recently~\cite{berg2018graph, ying2018graph, wang2019neural, zhou2022layer, zhou2022tale, zhou2023enhancing}.
The user-item interaction matrix naturally can be treated as a bipartite graph. GCN-based CF takes advantage of fusing both high-order information and the inherent graph structure.
GCNs are used to propagate information using the normalized adjacency matrix and aggregate information from neighbors via the nonlinear activation and linear transformation layers.
He \textit{et al.} \cite{he2020lightgcn} simplify the GCNs architecture by removing the feature transformation as well as nonlinear activation layers as they impose negative effect on recommendation performance.
In~\cite{chen2020revisiting}, the authors add a residual preference learning on GCN and obtain better recommendation performance.

\subsection{Self-supervised Learning}
SSL has achieved competitive results on various tasks in  vision and natural language processing domains. We review two lines of work on SSL.

\textbf{Contrastive learning}. 
Contrastive approaches learn representations by attracting the positive sample pairs and repulsing the negative sample pairs~\cite{hadsell2006dimensionality}.
A line of work~\cite{wu2018unsupervised, hjelm2018learning, henaff2020data, he2020momentum, chent2020simple, zhang2022diffusion, zhang2023multimodal} is developed based on this concept. 
These work benefits from a large number of negative samples, which require a memory bank~\cite{wu2018unsupervised} or a queue~\cite{he2020momentum} to store negative samples.
In~\cite{wu2021self}, the authors integrate supplemental signal into supervised baselines for contrastive learning, and show that it performs better than their baselines.
Following the line, Yu et al. propose a graph-augmentation free recommendation model~\cite{yu2022graph} to enforce the learning of uniform representations for users and items. The uniform representations can mitigate the popularity bias and achieve better recommendation accuracy.
Liu et al. summarize the contrastive learning applied on a broad fields, \eg NLP, Computer Vision, in~\cite{liu2021self}.

\begin{figure*}
	\centering
	\begin{subfigure}[b]{.45\linewidth}
		\centering
		\includegraphics[width=.85\linewidth]{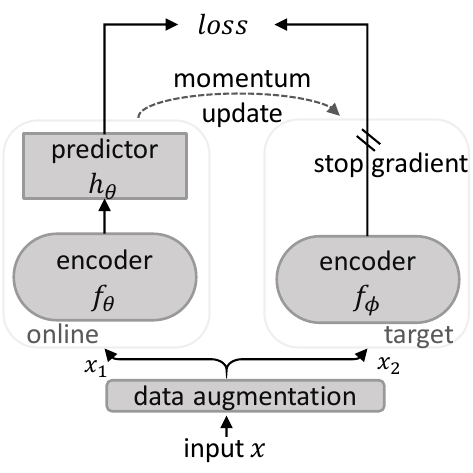}
		\caption{Overview of Siamese networks.}
		\label{fig:siam_neta}
	\end{subfigure}%
	\begin{subfigure}[b]{.45\linewidth}
		\centering
		\includegraphics[width=.85\linewidth]{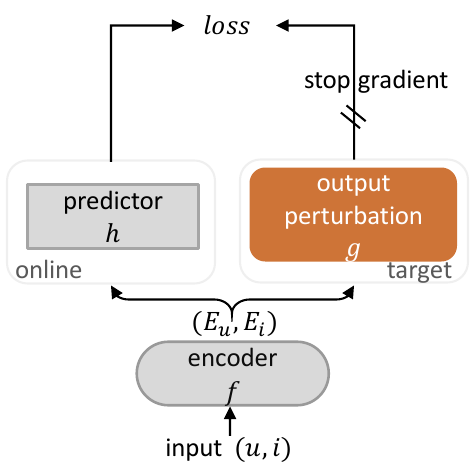}
		\caption{The SelfCF framework.}
		\label{fig:selfcf}
	\end{subfigure}%
	\caption{Comparison of Siamese network architectures. Input $x$ is an image. The input to SelfCF is the interaction pairs of users ($u$) and items ($i$).}
	\label{fig:siam_net}
\end{figure*}

\textbf{Siamese networks}. 
Siamese networks~\cite{bromley1993signature} are general models for comparing entities.
BYOL~\cite{grill2020bootstrap} and SimSiam~\cite{chen2021exploring} are two specializations of the Siamese network that achieve remarkable results by only using positive samples.
BYOL proposes two coupled networks (\textit{i.e.,} online and target networks) that are optimized and updated iteratively.
In detail, the online network is optimized towards the target network, while the target network is updated with a moving average of the online network to avoid collapse.
On the contrary, SimSiam verifies that a ``stop gradient'' operator is crucial in preventing collapse.
As a result, it removes the dashed ``momentum update'' line in Fig.~\ref{fig:siam_neta}.

Derived from BYOL, the recently proposed self-supervised framework, BUIR, learns the representation of users and items solely on positive interactions. It introduces different views by differentiating parameters of online and target networks.
However, the framework modifies the underlying logic of the encapsulated graph-based CF models for the sake of introducing contrastive user-item pairs.
In our solution, we choose to augment the output of encoder $f$ to generate two different but related embeddings for representation learning.
For comparison, we present our proposed framework specialized for CF, SelfCF, in Fig.~\ref{fig:selfcf}.
The framework shares the same encoder between online and target networks, thus reduces the unnecessary memory and computational resources for storing and executing an additional encoder of the target network.
We elaborate our framework in the following section.

\section{The SelfCF Framework}
Our framework~(shown in Fig.~\ref{fig:selfcf} ) partially inherits the Siamese network architecture of SimSiam, as shown in Fig.~\ref{fig:siam_net}. 
In our framework, SelfCF, the goal is to learn informative representations of users and items based on positive user-item interactions only.
The latent embeddings of users and items are learnt from the online network.
Analogous to convolutions~\cite{lecun1989backpropagation}, which is a successful inductive bias via weight-sharing for modeling translation-invariance, the weight-sharing Siamese networks can model invariance with regard to more complicated transformations ($e.g.$, data augmentations)~\cite{chen2021exploring}.
The online and target networks in SelfCF use a same copy of the parameters as well as the backbone for modeling representation invariance.
In addition, we drop the momentum encoder as used in BYOL and BUIR. 
As a result, with the same input, the online and target networks will generate the same output which makes the loss totally vanished.
We will discuss how to tackle this issue in the following section.

When considering data augmentations of input in CF, it is not a trivial task to distort the positive samples.
In vision domain, where SSL is popularly applied, images can be easily distorted under a wide range of transformations. 
However, positive user-item pairs are difficult to be distorted while preserving their representation invariance.
We use the following embedding perturbation techniques to achieve the same effect.
For reasons of clarity, we denote bold value $\textbf{E}$ as the embedding matrix of users and items within a batch, and differentiate the embedding matrix of users with $\textbf{E}_u$, vice visa.
The value $e$ in lowercase denotes the embedding of a user or item, specified as $e_u$ or $e_i$.

\begin{figure*}[t]
	\centering
	\includegraphics[width=0.96\columnwidth]{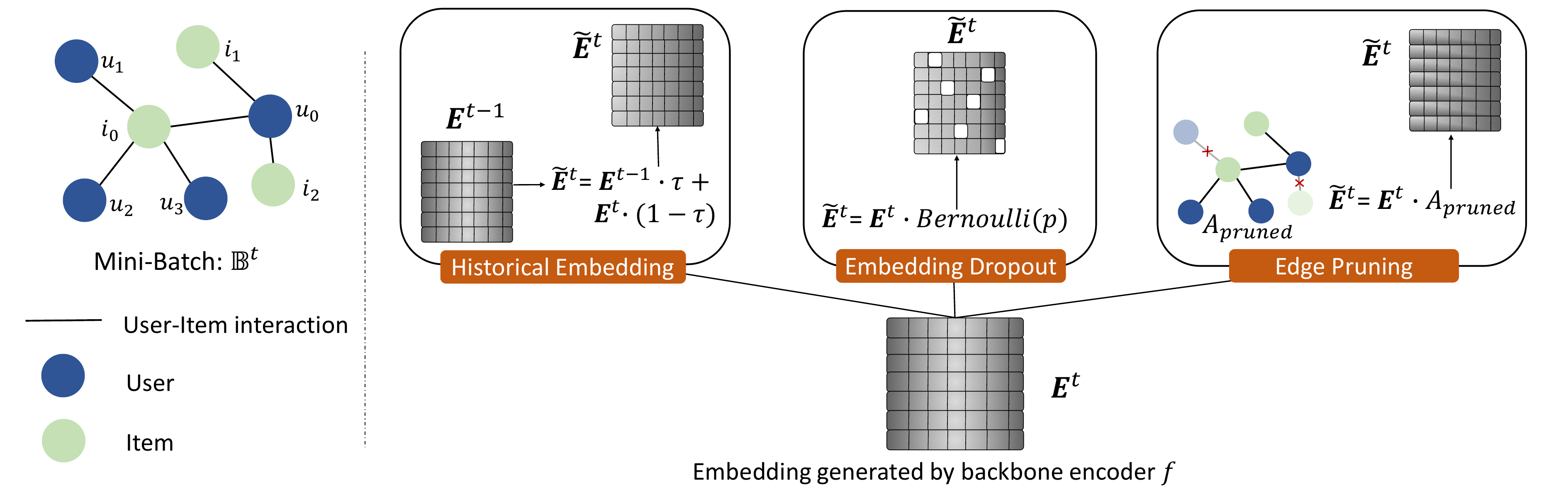}
	\caption{Illustration of output perturbation performed on a batch. The perturbed embedding is denoted as $\tilde{\textbf{E}}$.}
	\label{fig:pert}
\end{figure*}

\subsection{Data Augmentation via Output Perturbation}
In vision, researchers use image transformations to augment input data and generate two different but relative reviews. 
Instead, our framework augments the output embeddings of users and items to generate two contrastive views.
We propose three methods to introduce embedding perturbation in our framework, shown in Fig.~\ref{fig:pert}.
The historical embedding and embedding dropout are general techniques for output augmentation in our framework, while the edge pruning is specially designed for graph-based CF models.

\vspace{1.5ex}\noindent\textbf{Historical embedding.} 
We introduce embedding perturbation by utilizing historical embeddings~\cite{chen2017stochastic, fey2021gnnautoscale} from prior training iterations.
Specifically, we use a momentum update to generate the contrastive embeddings in the target network.
Suppose $\textbf{E}^t$ is the embeddings generated by a backbone encoder $f$ in a batch $\mathbb{B}^t$. 
The perturbed embeddings $\tilde{\textbf{E}}^t$ is calculated by combining of the output embeddings $\textbf{E}^t$ with the historical embedding $\textbf{E}^{t-1}$:
\begin{equation}
	\tilde{\textbf{E}}^t = \textbf{E}^{t-1} \tau + \textbf{E}^t (1 - \tau)
\end{equation}
where $\tau$ is a parameter controls the proportion of information preserved from a prior iteration. 

\vspace{1.5ex}\noindent\textbf{Embedding dropout.}
We apply the embedding dropout scheme to perturb the embeddings of users and items from the target network. 
In classical CF models, the parameters are not modified until the loss is backpropagated. 
With the same input, to avoid null loss resulted from these models, our framework adopts embedding dropout on the resulted users' and items' vectors, which is analogous to node dropout~\cite{srivastava2014dropout}.
In this way, the framework is able to generate two different but related views on the output, which are then feed into the loss function for optimization.
The resulted embeddings under a dropout ratio $p$ is calculated as:
\begin{equation}
	\tilde{\textbf{E}}^t = \textbf{E}^t \cdot Bernoulli(p)
\end{equation}

\vspace{1.5ex}\noindent\textbf{Edge pruning.}
As for graph-based CF models, the edge pruning method used in~\cite{rong2020dropedge, luo2021learning} provides an alternative way to augment the output embeddings.
With the user-item bipartite graph, we randomly prune a certain proportion of edges from the graph in each batch.
The output embeddings are updated by aggregating the embeddings of neighbors.
With the same positive user-item pair, the output is distorted with different adjacency matrix (neighbors). 
Let $A_{pruned}$ be the pruned adjacency matrix, then the resulted embeddings with edge pruning denote as:
\begin{equation}
	\tilde{\textbf{E}}^t = \textbf{E}^t \cdot A_{pruned}
\end{equation}
Note that, in implementation,  edge pruning would require to calculate the adjacency matrix of users and items, which is more expensive in computation than the embedding dropout technique.

To summarize, our framework augments the output via embedding perturbation in the target network instead of distorting the input directly as commonly used in vision domain.
It is worth noting that the historical embedding perturbation performs on embeddings from prior and current iteration, the embedding dropout perturbs the current embedding with noise, and the edge pruning method operates on future embeddings generated by stacking one more convolutional layer on current embeddings.
Both historical embedding perturbation and embedding dropout perturbation remove the requirements of auxiliary graphs to generate a contrastive view as in~\cite{wu2021self, lee2021bootstrapping, yu2022graph}.
We will discuss their performance with regard to this perspective in experiments section.

\subsection{The Loss Function}
Our framework, as shown in Fig.~\ref{fig:selfcf}, takes a positive user-item pair $(u,i)$ as input. 
The $(u,i)$ pair is initially processed by an encoder network $f$ in a backbone ($e.g.$ LightGCN~\cite{he2020lightgcn}).
The output of the encoder $f$ is then copied to the target network for embedding perturbation.
Formally, we denote the output of the encoder from the online network as $(e_u, e_i) = f(u, i)$. 
Finally, the linear predictor in our framework transforms the output $(e_u, e_i)$ with $(\dot{e}_u, \dot{e}_i) = h(e_u, e_i)$ and matches it to the perturbed embeddings $(\tilde{e}_u, \tilde{e}_i) = g(e_u, e_i)$ in other view like in BYOL~\cite{grill2020bootstrap} and SimSiam~\cite{chen2021exploring}.

We define a symmetrized loss function as the negative cosine similarity between $(\dot{e}_u, \tilde{e}_i)$ and $(\tilde{e}_u, \dot{e}_i)$:
\begin{equation}
	\mathcal{L} = \frac{1}{2} \mathcal{C}(\dot{e}_u, \tilde{e}_i) +  \frac{1}{2} \mathcal{C}(\tilde{e}_u, \dot{e}_i)
	\label{eq:loss}
\end{equation}
Function $\mathcal{C}(\cdot, \cdot)$ in the above equation is defined as:
\begin{equation}
	\mathcal{C}(e_u, e_i) = -\frac{(e_u)^T e_i}{||e_u||_2 ||e_i||_2},
\end{equation}
where $||\cdot||_2$ is $\ell_2$-norm.
The total loss is averaged over all user-item pairs in a batch.
The intuition behind this is that we intend to maximize the prediction of the perturbed item $i$ given a user $u$, and vice versa. The minimized possible value for this loss is $-1$.

Finally, we stop gradient on the target network and force the backpropagation of loss over the online network only.
We follow the stop gradient ($sg$) operator as in~\cite{grill2020bootstrap, chen2021exploring}, and implement the operator by updating Equation~\ref{eq:loss} as:
\begin{equation}
	\mathcal{L} = \frac{1}{2} \left( \mathcal{C}(\dot{e}_u, sg(\tilde{e}_i)) + \mathcal{C}(sg(\tilde{e}_u), \dot{e}_i) \right).
	\label{eq:loss_final}
\end{equation}
With the stop gradient operator, the target network receives no gradient from $(\tilde{e}_u, \tilde{e}_i)$.
However, the encoder $f$ in the online network receives gradients from user-item pair $(\dot{e}_u, \dot{e}_i)$, and optimizes its parameters towards the global optimum. 
Conversely, the removal of this operator can cause instability in online network learning, which we will verify this through ablation study. The reason is that the online and target networks simulate the student-teacher-like network~\cite{tarvainen2017mean} in which only the online network is optimized to predict the positively interacted item (user) presented by the target network.
Additionally, we add regularization penalty on the online embeddings $(\ie e_u$ and  $e_i)$ and the predictor $h$. The final loss function is: 
\begin{equation}
	\mathcal{L} = \frac{1}{2} \left( \mathcal{C}(\dot{e}_u, sg(\tilde{e}_i)) + \mathcal{C}(sg(\tilde{e}_u), \dot{e}_i) \right) + \lambda_1 \cdot (||e_u||^2_2 + ||e_i||^2_2) + \lambda_2 \cdot (||h||^2_1),
	\label{eq:loss_final_reg}
\end{equation}
where $||\cdot||_1$ is $\ell_1$-norm.
The pseudo-code of SelfCF is in Algorithm 1.

\begin{algorithm}
	\caption{PyTorch-style pseudo-code for SelfCF.}
	\label{influx}
	\begin{algorithmic}[1]
		\Require user-item interaction set $\mathbb{B}$
		\Require $f, h, g$ \Comment{encoder, predictor, output perturbation}
		\For{$\mathbb{B}^t ~ in ~ \mathbb{B}$}\Comment{load a batch} 
		\State $(\textbf{E}^t_u, \textbf{E}^t_i) = f(\mathbb{B}^t)$ \Comment{output of encoder} 
		\State $(\dot{\textbf{E}}^t_u, \dot{\textbf{E}}^t_i) = h(\textbf{E}^t_u, \textbf{E}^t_i)$ \Comment{output of predictor} 
		\State $(\tilde{\textbf{E}}^t_u, \tilde{\textbf{E}}^t_i) = g(\textbf{E}^t_u, \textbf{E}^t_i)$ \Comment{output of perturbation} 
		\State $\mathcal{L} = \frac{1}{2} \left( \mathcal{C}(\dot{\textbf{E}}^t_u, sg(\tilde{\textbf{E}}^t_i)) + \mathcal{C}(sg(\tilde{\textbf{E}}^t_u), \dot{\textbf{E}}^t_i) \right) + \lambda_1 \cdot (||\textbf{E}^t_u||^2_2 + ||\textbf{E}^t_i||^2_2) + \lambda_2 \cdot (||h||^2_1)$ \Comment{Eq. 7}
		\State $\mathcal{L}$.backward() \Comment{back-propagate}
		\State update($f, h$) \Comment{parameters update}
		\EndFor
		\State
		\State \textbf{def} predict($e_u, e_i$):
		\Comment{calculate recommendation score}
		\State \qquad return $s(e_u, e_i)$
		\Comment{Eq. 8}
	\end{algorithmic}
\end{algorithm}

\subsection{Top-$K$ Recommendation}
Classical CF methods recommend top-$K$ items by ranking scores of the inner product of a user embedding with all candidate item embeddings.
However, in SSL, we minimize the predicted loss between $u$ and $i$ for each positive interaction $(u, i)$.
Intuitively, we predict the future interaction score based on a cross-prediction task~\cite{lee2021bootstrapping}. 
That is, we both predict the interaction probability of item $i$ with $u$ and the probability of user $u$ with $i$.
Given $(e_u, e_i)$ being the output of the encoder $f$, the recommendation score is calculated as:
\begin{equation}
	s(e_u, e_i) = h(e_u) \cdot (e_i)^T + e_u \cdot h(e_i)^T
\end{equation}
It is worth noting that since the encoder $f$ is shared between both online and target networks, we use the representations obtained from the online network to predict top-$K$ items for each user.

\section{Experiments}
We evaluate the framework on four publicly available datasets and compare its performance with BUIR~\cite{lee2021bootstrapping} and eight baselines by encapsulating BPR and LightGCN as our backbone.
Our framework is mainly compared with BUIR, as it is the only recommendation framework that works without negative samples.
All baselines as well as our frameworks are trained on a single GeForce RTX 2080 Ti (11 GB).

We list the research questions addressed in our evaluation as follows:
\begin{description}
	\item{\textbf{RQ1}}: Whether the self-supervised models that only leverage positive user-item interactions can outperform their supervised counterparts?
	\item{\textbf{RQ2}}: How SelfCF shapes the recommendation results for cold-start and loyal users? 
	\item{\textbf{RQ3}}: Why SelfCF works, and which component is essential in preventing collapsing?
\end{description}
We address the first research question by evaluating our framework against supervised baselines with four datasets under six evaluation metrics.
Next, we dive into the recommendation results of the baselines under both supervised and self-supervised settings and analyze their performance on users with different number of interactions.
Finally, to investigate the underlying mechanisms of SelfCF, we perform ablation study on the components of SelfCF, such as the linear predictor, the loss function \textit{etc.}

\begin{table}
	\centering
	\caption{Statistics of the experimented data. }
	\label{tab:dataset}
	\begin{tabular}{lrrrr}
		\toprule
		Dataset & \# of Users & \# of Items & \# of Interactions & Sparsity \\
		\midrule
		Arts & 45,624 & 21,104 & 396,556 & 99.9588\%\\
		Games & 50,677 & 16,897 & 454,529 & 99.9469\%\\
		Food & 115,144 & 39,688 & 1,025,169 & 99.9776\%\\
		COCO & 144,773 & 20,969 & 1,204,697 & 99.9603\%\\
		\bottomrule
	\end{tabular}
\end{table}

\subsection{Dataset Description.} 
We choose the evaluated datasets carefully by considering the following principles in order to introduce as much as diversity.
\begin{description}
	\item{\textbf{Domain}}: Interactions within the same domain may exhibit similar patterns across datasets. Hence, we choose evaluation datasets from two different domains ranging from education to e-commerce under different categories.
	\item{\textbf{Released date}}: Existing recommender systems usually evaluated on out-dated datasets nearly collected 10 years ago. With the rapid growth of e-commerce platforms, user behaviors are gradually shaped with online purchasing. 
	\item{\textbf{Graph size}}: The user-item interactions can be viewed as a bipartite graph (Fig.~\ref{fig:uigraph}), we consider the graph size with the number of nodes ranging from 10K to 100K. 
\end{description}
We describe each dataset with regard to the above selection principles.
\begin{itemize}
	\item Amazon Video Games (\textit{Games}): This is a newly released version of the Amazon-Video-Games review dataset in 2018. We select the rating only version for evaluation. Dataset is available from~\cite{ni2019justifying}~\footnote{https://nijianmo.github.io/amazon/index.html}.
	\item Amazon Arts, Crafts and Sewing (\textit{Arts}): This dataset is similar to the Amazon Video Games dataset under a different genre.
	\item Amazon Grocery and Gourmet Food (\textit{Food}): This dataset has a large-scale interaction graph with more than 100K users. 
	\item COCO: A large-scale dataset from education domain. The raw dataset includes over 43K
	online courses and 2.5M learners~\cite{dessi2018coco}.
\end{itemize}
All raw datasets are preprocessed with a 5-core setting on both items and users and the filtered results are presented in Table~\ref{tab:dataset}. 

\subsection{Encapsulated Baselines and Framework BUIR.}

To compare the performance of our proposed framework, we first consider the following baselines that adopt negative sampling for supervised learning except the popularity algorithm.
\begin{itemize}
	\item Pop: Popularity algorithm recommends the most popular items to each user.
	\item BPR~\cite{rendle2009bpr}: A matrix factorization model optimized by a pairwise ranking loss in a Bayesian way.
	\item MultiVAE~\cite{liang2018variational}: It is a generative model that adopts variational auto-encoder (VAE) for item-based CF.
	It uses a multinomial likelihood to fit the distribution of data and adopts Bayesian inference for parameter estimation. 
	\item EHCF~\cite{chen2020efficienth}: 
	This is an efficient recommendation model that learns the representations of users and items by reconstructing the interaction matrix without negative sampling. 
	It takes all unobserved user-item pairs as negative samples.
	\item NGCF~\cite{wang2019neural}: 
	This model explicitly injects collaborative signal from high-order connectivity of user-item graph into the embedding learning process.
	\item LR-GCCF~\cite{chen2020revisiting}: The model first simplifies the vanilla GCN by removing nonlinear function, then it uses a residual preference learning process for prediction.
	\item LightGCN~\cite{he2020lightgcn}: This is a simplified graph convolution network that only performs linear propagation and aggregation between neighbors. The hidden layer embeddings are averaged to calculate the final user/item embeddings for prediction.
	\item SimGCL~\cite{yu2022graph}: This self-supervised model injects uniform noises into the latent embeddings to generate contrastive views.
\end{itemize}

We also consider the following self-supervised frameworks that learn the representations of users and items without negative samples.
Our framework is mainly compared with BUIR~\cite{lee2021bootstrapping}, a self-supervised framework that is derived from BYOL~\cite{grill2020bootstrap}. Its architecture follows the Siamese network in Fig.~\ref{fig:siam_neta}.
To compare the performance of our proposed framework, we encapsulate two state-of-the-art models, BPR and LightGCN, into the frameworks. That is, we substitute the encoder $f$ in Fig.~\ref{fig:selfcf} with BPR and LightGCN, respectively.

\begin{itemize}
	\item BUIR~\cite{lee2021bootstrapping}: This framework uses asymmetric network architecture to update its backbone network parameters.
	\item SelfCF$_{\textrm{he}}$: 
	Our proposed framework that uses \textbf{historical embedding} for data augmentation. 
	\item SelfCF$_{\textrm{ed}}$: 
	Our proposed framework that uses \textbf{embedding dropout} for data augmentation. 
	\item SelfCF$_{\textrm{ep}}$: 
	Our proposed framework that uses \textbf{edge pruning} for data augmentation. 
\end{itemize}
To demonstrate the generalization of our framework, we consider two backbone networks for BUIR and SelfCF$_{\textrm{ed}}$, the classic BPR and recently graph-based LightGCN. Other frameworks will only use LightGCN as their backbone network because LightGCN always shows better performance against BPR.

\subsection{Evaluation Metrics.}
We use $Recall@K$ and $NDCG@K$ computed by the all-ranking protocol as the evaluation metrics for recommendation accuracy comparison.
In the recommendation phase, all items that have not been interacted with a specific user are regarded as candidates. That is, we do not use sampled evaluation.

Formally, we define $I^r_u(i)$ as the $i$-th ranked item recommended for $u$, $\mathcal{I}[\cdot]$ is the indicator function, and $I^t_u$ is the set of items that user $u$ interacted in the testing data.

Recall@K for users $u$ is:
\begin{equation}
	Recall@K(u) = \frac{\sum^K_{i=1} \mathcal{I}[I^r_u(i) \in I^t_u]}{|I^t_u|}
\end{equation}

The Discounted Cumulative Gain (DCG@K) is:
\begin{equation}
	DCG@K(u) = \sum^K_{i=1} \frac{2^{\mathcal{I}[I^r_u(i) \in I^t_u]}-1}{log(i+1)}
\end{equation}
$NDCG@K$ is normalized to [0, 1] with $NDCG@K=DCG@K/IDCG@K$, where $IDCG@K$ is calculated by sorting the interacted items in the testing data at top and then use the formula for $DCG@K$.
We set $K=10$, $K=20$ and $K=50$ in our experimental comparison, respectively. 
For simplicity, we denote $Recall@K$ and $NDCG@K$ with $R@K$ and $N@K$ in the following sections.

\begin{table}
	\centering
	\caption{Hyper-parameter exploration. }
	\label{tab:selfcf_hyper}
	\setlength\tabcolsep{6.0pt} 
	\begin{tabular}{llcr}
		\toprule
		\textbf{Framework} & \textbf{Backbone model} & \textbf{Para.} & \textbf{Tuning Range} \\
		\midrule
		\multirow{4}{*}{\shortstack{SelfCF$_{\textrm{he}}$ \\ \& \\ SelfCF$_{\textrm{ep}}$}}& \multirow{4}{*}{LightGCN} & \textit{layers} & [1, 2, 3, 4] \\
		&  & $momentum$ & [0.1, 0.2, 0.5] \\
		&  & $\lambda_1$ & [0.0, 1e-01, 1e-02, 1e-03, 1e-04, 1e-05] \\
		&  & $\lambda_2$ & [0.0] \\
		\midrule
		\multirow{7}{*}{SelfCF$_{\textrm{ed}}$} & \multirow{3}{*}{BPR} & \textit{dropout} & [0.05] \\
		&  & $\lambda_1$ & [0.0] \\
		&  & $\lambda_2$ & [1e-02] \\
		\cmidrule{2-4}
		& \multirow{4}{*}{LightGCN} & \textit{layers} & [1, 2] \\
		&  & \textit{dropout} & [0.1, 0.2, 0.5] \\
		&  & $\lambda_1$ & [0.0, 1e-01, 1e-02, 1e-03, 1e-04, 1e-05] \\
		&  & $\lambda_2$ & [0.0] \\
		\bottomrule
	\end{tabular}
\end{table}

\subsection{Hyper-parameters Settings.}
Same as other work~\cite{chen2020revisiting,he2020lightgcn}, we fix the embedding size of both users and items to 64 for all models, initialize the embedding parameters with the Xavier method~\cite{glorot2010understanding}, and use Adam~\cite{kingma2015adam} as the optimizer.
For a fair comparison, we carefully tune the parameters of each model following their published papers.
For our proposed frameworks, we perform grid search across all datasets to conform the optimal settings. We summarize the settings in Table~\ref{tab:selfcf_hyper}. We penalize the predictor with $L_1$ regularization when BPR is encapsulated, otherwise, we use $L_2$ regularization. The reason is that BPR learns the embeddings of users and items without leveraging graph structure and opts to over-fitting. We add $L_1$ regularization to learn a sparsified predictor.
For convergence consideration, the early stopping and total epochs are fixed at 50 and 1000, respectively. 
Following~\cite{wang2019neural}, we use Recall@20 on validation data as the training stopping indicator.
We implement our model on top of Recbole~\cite{zhao2021recbole} at: https://github.com/enoche/SelfCF.

\begin{table*}
	\caption{Overall performance comparison. We mark the global best results on each dataset under each metric in \textbf{boldface}, and the second best \underline{underlined}.
		We also calculate the performance improvement by SelfCF on BUIR over each evaluation metric as $\Delta$. ``NS'' denotes Negative Samples.}
	\label{tab:performance}
	\centering
	\def\arraystretch{0.95}%
	\setlength\tabcolsep{3.6pt} 
	\begin{tabular}{lllcccccc}
		\toprule
		\textit{\textbf{Arts}} & \textbf{Framework} & \textbf{Model} & \textbf{R@10} & \textbf{R@20} & \textbf{R@50} & \textbf{N@10} & \textbf{N@20} & \textbf{N@50} \\ 
		\midrule
		\textbf{Non-parametric} &- & Pop & 0.0091 & 0.0164 & 0.0283 & 0.0072 &  0.0095 & 0.0128 \\
		\midrule
		\multirow{6}{*}{\begin{tabular}{@{}l@{}}\textbf{Supervised}  \\ (with NS) \end{tabular}} & \multirow{6}{*}{-} & BPR & 0.0201 & 0.0327 & 0.0589  & 0.0137 & 0.0177 & 0.0245 \\
		& & MultiVAE &  0.0171 & 0.0268 & 0.0503 & 0.0113 & 0.0145 & 0.0205  \\
		\textbf{Supervised} &  & EHCF & 0.0202 & 0.0319 & 0.0567  & 0.0136 & 0.0175 & 0.0240 \\
		(with NS) &  & NGCF &  0.0205 & 0.0342 & 0.0623 & 0.0142 & 0.0186 & 0.0260  \\
		& & LR-GCCF & 0.0221 & 0.0365 & 0.0636 & 0.0151 & 0.0197 & 0.0268 \\
		& & LightGCN & 0.0231 & 0.0371 & 0.0663 & 0.0156 & 0.0201 & 0.0277 \\	
		\midrule	
		\begin{tabular}{@{}l@{}}\textbf{Self-Supervised}  \\ (with NS) \end{tabular} &  & SimGCL & 0.0198 & 0.0322 & 0.0558 & 0.0133 & 0.0172 & 0.0234 \\
		\midrule
		\multirow{10}{*}{\begin{tabular}{@{}l@{}}\textbf{Self-Supervised}  \\ (without NS) \end{tabular}} & \multirow{2}{*}{BUIR} & BPR & 0.0197 & 0.0309 & 0.0560 & 0.0139 & 0.0174 & 0.0239  \\
		&  & LightGCN & 0.0208 & 0.0334 & 0.0636 & 0.0149 & 0.0190 & 0.0270  \\
		\cmidrule{2-9}
		& \multirow{2}{*}{SelfCF$_{\textrm{he}}$} & LightGCN & 0.0236 & \textbf{0.0397} & \underline{0.0709} & 0.0157 & \underline{0.0208} & 0.0289 \\
		&  & $\Delta$ & 13.46\% & 18.86\% & 11.48\% & 5.37\% & 9.47\% & 7.04\%  \\
		\cmidrule{2-9}
		& \multirow{4}{*}{SelfCF$_{\textrm{ed}}$} & BPR & 0.0231 & 0.0354 & 0.0632 & 0.0157 & 0.0197 & 0.0269 \\
		& & $\Delta$ & 17.26\% & 14.56\% & 12.86\% & 12.95\% & 13.22\% & 12.55\%  \\
		& & LightGCN & \textbf{0.0246} & 0.0391 & 0.0708 & \textbf{0.0170} & \textbf{0.0218} & \textbf{0.0300} \\
		& & $\Delta$ & 18.27\% & 17.07\% & 11.32\% & 14.09\% & 14.74\% & 11.11\%  \\
		\cmidrule{2-9}
		& \multirow{2}{*}{SelfCF$_{\textrm{ep}}$} & LightGCN & \underline{0.0239} & \underline{0.0395} & \textbf{0.0714} & \underline{0.0158} & \underline{0.0208} & \underline{0.0290} \\
		& & $\Delta$ & 14.90\% & 18.26\% & 12.26\% & 6.04\% & 9.47\% & 7.41\%  \\
		\bottomrule
		\bottomrule
		\textit{\textbf{Games}} &  \textbf{Framework} & \textbf{Model} & \textbf{R@10} & \textbf{R@20} & \textbf{R@50} & \textbf{N@10} & \textbf{N@20} & \textbf{N@50} \\ 
		\midrule
		\textbf{Non-parametric} &- & Pop & 0.0117 & 0.0175 & 0.0379 & 0.0049 &  0.0067 & 0.0117 \\
		\midrule
		\multirow{6}{*}{\begin{tabular}{@{}l@{}}\textbf{Supervised}  \\ (with NS) \end{tabular}} & \multirow{6}{*}{-} & BPR & 0.0210 & 0.0369 & 0.0699 & 0.0135 &  0.0183 & 0.0265 \\
		& & MultiVAE &  0.0238 & 0.0376 & 0.0718 & 0.0154 & 0.0196 & 0.0280  \\
		&  & EHCF &  0.0278 & 0.0445 & 0.0772 & 0.0175 & 0.0227 & 0.0308 \\
		& & NGCF &  0.0254 & 0.0425 & 0.0825 & 0.0166 & 0.0217 & 0.0314  \\
		& & LR-GCCF & 0.0259 & 0.0446 & 0.0824 & 0.0171 & 0.0228 & 0.0320 \\
		& & LightGCN & 0.0275 & 0.0461 & 0.0841 & 0.0175 & 0.0231 & 0.0326 \\
		\midrule
		\begin{tabular}{@{}l@{}}\textbf{Self-Supervised}  \\ (with NS) \end{tabular} &  & SimGCL & \textbf{0.0310} & \underline{0.0502} & \underline{0.0879} & \textbf{0.0194} & \underline{0.0251} & \underline{0.0344} \\
		\midrule
		\multirow{10}{*}{\begin{tabular}{@{}l@{}}\textbf{Self-Supervised}  \\ (without NS) \end{tabular}} & \multirow{2}{*}{BUIR} & BPR & 0.0217 & 0.0361 & 0.0674 & 0.0135 & 0.0180 & 0.0257 \\
		&  & LightGCN & 0.0227 & 0.0384 & 0.0749 & 0.0143 & 0.0192 & 0.0282 \\
		\cmidrule{2-9}
		& \multirow{2}{*}{SelfCF$_{\textrm{he}}$} & LightGCN & 0.0295 & 0.0473 & 0.0859 & 0.0187 & 0.0241 & 0.0336 \\
		&  & $\Delta$ & 29.96\% & 23.18\% & 14.69\% & 30.77\% & 25.52\% & 19.15\% \\
		\cmidrule{2-9}
		& \multirow{4}{*}{SelfCF$_{\textrm{ed}}$} & BPR & 0.0241 & 0.0402 & 0.0744 & 0.0152 & 0.0200 & 0.0285 \\
		& & $\Delta$ & 11.06\% & 11.36\% & 10.39\% & 12.59\% & 11.11\% & 10.89\%  \\
		& & LightGCN & 0.0289 & 0.0485 & 0.0857 & 0.0181 & 0.0240 & 0.0332 \\
		& & $\Delta$ & 27.31\% & 26.30\% & 14.42\% & 26.57\% & 25.00\% & 17.73\% \\
		\cmidrule{2-9}
		& \multirow{2}{*}{SelfCF$_{\textrm{ep}}$} & LightGCN & \underline{0.0301} & \textbf{0.0517} & \textbf{0.0930} & \underline{0.0189} & \textbf{0.0255} & \textbf{0.0358} \\
		& & $\Delta$ & 32.60\% & 34.64\% & 24.17\% & 32.17\% & 32.81\% & 26.95\% \\
		\bottomrule
	\end{tabular}
\end{table*}

\begin{table*}
	\centering
	\def\arraystretch{0.95}%
	\setlength\tabcolsep{3.6pt} 
	\begin{tabular}{lllcccccc}
		\toprule
		\bottomrule
		\textit{\textbf{Food}} & \textbf{Framework} & \textbf{Model}  & \textbf{R@10} & \textbf{R@20} & \textbf{R@50} & \textbf{N@10} & \textbf{N@20} & \textbf{N@50} \\ 
		\midrule
		\textbf{Non-parametric} &- & Pop & 0.0125 & 0.0189 & 0.0346 & 0.0112 &  0.0133 & 0.0173 \\
		\midrule
		\multirow{6}{*}{\begin{tabular}{@{}l@{}}\textbf{Supervised}  \\ (with NS) \end{tabular}} & \multirow{6}{*}{-} & BPR & 0.0138 & 0.0222 & 0.0390 & 0.0097 & 0.0124 & 0.0167 \\
		& & MultiVAE &  0.0133 & 0.0208 & 0.0374 & 0.0092 & 0.0116 & 0.0159  \\
		&  & EHCF & 0.0158 & 0.0243 & 0.0416  & 0.0111 & 0.0137 & 0.0182 \\
		&  & NGCF &  0.0158 & 0.0254 & 0.0456 & 0.0102 & 0.0132 & 0.0185  \\
		& & LR-GCCF & 0.0172 & 0.0277 & 0.0478 & 0.0120 & 0.0154 & 0.0206  \\
		& & LightGCN & 0.0184 & 0.0286 & 0.0497 & 0.0125 & 0.0157 & 0.0211  \\	
		\midrule	
		\begin{tabular}{@{}l@{}}\textbf{Self-Supervised}  \\ (with NS) \end{tabular} & - 
		& SimGCL & 0.0173 & 0.0265 & 0.0453 & 0.0116 & 0.0147 & 0.0195 \\
		\midrule
		\multirow{10}{*}{\begin{tabular}{@{}l@{}}\textbf{Self-Supervised}  \\ (without NS) \end{tabular}} & \multirow{2}{*}{BUIR} & BPR & 0.0113 & 0.0178 & 0.0313 & 0.0075 & 0.0096 & 0.0130  \\
		&  & LightGCN & 0.0145 & 0.0236 & 0.0469 & 0.0111 & 0.0141 & 0.0201  \\
		\cmidrule{2-9}
		& \multirow{2}{*}{SelfCF$_{\textrm{he}}$} & LightGCN & \underline{0.0195} & \underline{0.0299} & \underline{0.0516} & \underline{0.0132} & \underline{0.0166} & \underline{0.0221} \\
		&  & $\Delta$ & 34.48\% & 26.69\% & 10.02\% & 18.92\% & 17.73\% & 9.95\%  \\
		\cmidrule{2-9}
		& \multirow{4}{*}{SelfCF$_{\textrm{ed}}$} & BPR & 0.0165 & 0.0259 & 0.0443 & 0.0111 & 0.0141  & 0.0188 \\
		& & $\Delta$ & 46.02\% & 45.51\% & 41.53\% & 48.00\% & 46.88\% & 44.62\% \\
		& & LightGCN & \textbf{0.0198} & \textbf{0.0316} & \textbf{0.0555} & \textbf{0.0135} & \textbf{0.0173} & \textbf{0.0235} \\
		& & $\Delta$ & 36.55\% & 33.90\% & 18.34\% & 21.62\% & 22.70\% & 16.92\%  \\
		\cmidrule{2-9}
		& \multirow{2}{*}{SelfCF$_{\textrm{ep}}$} & LightGCN & 0.0186 & 0.0296 & 0.0514 & 0.0126 & 0.0161 & 0.0216 \\
		& & $\Delta$ & 28.28\% & 25.42\% & 9.59\% & 13.51\% & 14.18\% & 7.46\%  \\
		\bottomrule		
		\textit{\textbf{COCO}} & \textbf{Framework} & \textbf{Model} &\textbf{R@10} & \textbf{R@20} & \textbf{R@50} & \textbf{N@10} & \textbf{N@20} & \textbf{N@50} \\ 
		\midrule
		\textbf{Non-parametric} &- & Pop & 0.0574 & 0.0798 & 0.1393  & 0.0318 & 0.0385 & 0.0525 \\
		\midrule
		\multirow{6}{*}{\begin{tabular}{@{}l@{}}\textbf{Supervised}  \\ (with NS) \end{tabular}} & \multirow{6}{*}{-} & BPR & 0.1181 & 0.1745 & 0.2681  & 0.0741 & 0.0908 & 0.1129 \\
		&  & MultiVAE & \underline{0.1243} & 0.1816 & 0.2786  & 0.0775 & \underline{0.0946} & \underline{0.1175} \\
		\textbf{Supervised} &  & EHCF & 0.1146 & 0.1674 & 0.2507  & 0.0724 & 0.0880 & 0.1078 \\
		(with NS) &  & NGCF & 0.1210 & \underline{0.1817} & \underline{0.2843}  & 0.0740 & 0.0921 & 0.1163 \\
		&  & LR-GCCF & 0.1215 & 0.1784 & 0.2734 & 0.0754 & 0.0923 & 0.1147 \\
		&  & LightGCN & 0.1213 & 0.1781 & 0.2723 & 0.0762 & 0.0932 & 0.1154 \\			
		\midrule	
		\begin{tabular}{@{}l@{}}\textbf{Self-Supervised}  \\ (with NS) \end{tabular} & - & SimGCL & 0.1238 & 0.1758 & 0.2564 & \underline{0.0784} & 0.0939 & 0.1130 \\
		\midrule
		\multirow{10}{*}{\begin{tabular}{@{}l@{}}\textbf{Self-Supervised}  \\ (without NS) \end{tabular}} & \multirow{2}{*}{BUIR} & BPR & 0.0977 & 0.1445 & 0.2222 & 0.0601 & 0.0740 & 0.0924  \\
		&  & LightGCN & 0.1162 & 0.1745 & 0.2672 & 0.0719 & 0.0893 & 0.1113  \\
		\cmidrule{2-9}
		& \multirow{2}{*}{SelfCF$_{\textrm{he}}$} & LightGCN & 0.1147 & 0.1758 & 0.2722 & 0.0716 & 0.0898 & 0.1127 \\
		&  & $\Delta$ & -1.29\% & 0.74\% & 1.87\% & -0.42\% & 0.56\% & 1.26\%  \\
		\cmidrule{2-9}
		& \multirow{4}{*}{SelfCF$_{\textrm{ed}}$} & BPR & 0.1126 & 0.1672 & 0.2508 & 0.0684 & 0.0847 & 0.1046 \\
		& & $\Delta$ & 15.25\% & 15.71\% & 12.87\% & 13.81\% & 14.46\% & 13.20\%  \\
		& & LightGCN & \textbf{0.1287} & \textbf{0.1892} & \textbf{0.2877} & \textbf{0.0796} & \textbf{0.0977} & \textbf{0.1210} \\
		& & $\Delta$ & 10.76\% & 8.42\% & 7.67\% & 10.71\% & 9.41\% & 8.72\%  \\
		\cmidrule{2-9}
		& \multirow{2}{*}{SelfCF$_{\textrm{ep}}$} & LightGCN & 0.1174 & 0.1734 & 0.2712 & 0.0740 & 0.0906 & 0.1137 \\
		& & $\Delta$ & 1.03\% & -0.63\% & 1.50\% & 2.92\% & 1.46\% & 2.16\%  \\
		\bottomrule
	\end{tabular}
\end{table*}

\subsection{Overall Comparison}
While we acknowledge the significance of online evaluation for recommender systems, it is not feasible to evaluate our model in such a manner in an academic environment. 
Therefore, to avoid data leakage under offline evaluation~\cite{sun2022counter}, we adopt the evaluation setting used in~\cite{li2020dynamic, chen2021modeling}, which involves splitting the data chronologically in a 7:1:2 ratio for training, validation, and testing.
We define the global comparison perspective as the comparison across supervised and self-supervised baselines, while the local comparison perspective as the comparison between self-supervised frameworks BUIR and SelfCF.
We analyze the comparison results with regard to recommendation accuracy (Table~\ref{tab:performance}) under the following perspectives:
\begin{itemize}
	\item \textbf{Classic CF vs. Graph-based CF.} 
	In general, graph-based CF (\ie NGCF, LR-GCCF, LightGCN) performs better than other supervised baselines.
	We speculate the graph-based CF model naturally encodes structural embedding that is preferred for contrastive learning. 
	Analogously, self-supervised frameworks encapsulated with LightGCN also have better performance. 
	The performance of LightGCN under SelfCF is on par or better than that of under supervised learning.
	Classic CF models, $e.g.$ BPR, use pairwise learning to differentiate positive and negative user-item samples which encode less information between positive instances, resulting in a worse performance under the self-supervised framework, BUIR. 
	On the contrary, in our framework, we penalize the predictor $h$ with $L1$ regularization term. As a result, a sparse and weak predictor can encourage the framework to learn informative representations for users and items.  
	\item \textbf{Comparison between self-supervised frameworks.} When compared between frameworks without negative samples, our proposed framework SelfCF$_{\textrm{ed}}$ improves BUIR on every evaluation metric across all datasets.
	The proposed framework with three output perturbations takes significant improvement, as high as 17.79\% over four datasets on average. In particular, our framework SelfCF$_{\textrm{ed}}$ gains 21.19\% over BUIR when both use BPR as the backbone network. It is worth mentioning that SimGCL leveraging negative samples for representation learning obtains competitive performance on ranking metric (\eg NDCG@10).
	\item \textbf{Output perturbation techniques in SelfCF.}
	Among the three output perturbation techniques, history embedding technique integrates the embedding from previous training iteration; embedding dropout technique introduces noise on the current output embedding; and edge pruning technique achieves embedding augmentation by merging embedding from neighbors. Within the three proposed output perturbation techniques, Table~\ref{tab:performance} shows the embedding dropout technique is preferable across all datasets. 
	The reason is that the embedding dropout and edge pruning techniques can remove the noise information and preserve the salient features in the embeddings. 
	However, the embedding dropout is better than the edge pruning technique in retaining the similarity between the original embedding and the augmented embedding.
\end{itemize}

We conclude our analysis to address research question \textbf{RQ1}: Both classical CF and graph-based model CF can benefit from SelfCF. Specially, the supervised counterparts, BPR and LightGCN, can be improved with up to 7.36\% and 6.55\% across the four datasets under SelfCF$_{\textrm{ed}}$, respectively.

\subsection{Efficiency of SelfCF}
We evaluate the efficiency of SelfCF compared with LightGCN with regard to the number of layers in Table~\ref{tab:layer-comp}.
From the results, we observe SelfCF$_{\textrm{ed}}$ is on par or better than LightGCN stacked with 4-layer, but requires only one half to one quarter training time of LightGCN.

\begin{table}
	\caption{Efficiency of SelfCF.}
	\label{tab:layer-comp}
	\setlength\tabcolsep{4.0pt} 
	\begin{tabular}{llccccccc}
		\toprule
		\textbf{Dataset} & \textbf{Model} & \textbf{R@10} & \textbf{R@20} & \textbf{R@50} & \textbf{N@10} & \textbf{N@20} & \textbf{N@50} & \textbf{Time} ($s$)\\
		\midrule
		\multirow{6}{*}{\textit{\textbf{Games}}} & {SelfCF$_{\textrm{ed}}$} 1-Layer & 0.0274 & 0.0456 & 0.0857 & 0.0175 & 0.0231 & 0.0332 & 3.19\\
		& {SelfCF$_{\textrm{ed}}$} 2-Layer & 0.0289 & 0.0485 & 0.0857 & 0.0181 & 0.0240 & 0.0332 & 3.75\\
		\cmidrule{2-9}
		& LightGCN 4-Layer & 0.0275 & 0.0461 & 0.0841 & 0.0175 & 0.0231 & 0.0326 & 8.22\\
		& LightGCN 3-Layer & 0.0270 & 0.0458 & 0.0836 & 0.0176 & 0.0233 & 0.0326 & 7.60\\
		& LightGCN 2-Layer & 0.0271 & 0.0454 & 0.0818 & 0.0174 & 0.0230 & 0.0320 & 6.78\\
		& LightGCN 1-Layer & 0.0263 & 0.0448 & 0.0798 & 0.0172 & 0.0228 & 0.0315 & 5.05\\
		\midrule
		\multirow{6}{*}{\textit{\textbf{Food}}} & {SelfCF$_{\textrm{ed}}$} 1-Layer & 0.0197 & 0.0316 & 0.0547 & 0.0135 & 0.0173 & 0.0233 & 14.09\\
		& {SelfCF$_{\textrm{ed}}$} 2-Layer & 0.0198 & 0.0316 & 0.0555 & 0.0135 & 0.0173 & 0.0235 & 17.20\\
		\cmidrule{2-9}
		& LightGCN 4-Layer & 0.0184 & 0.0286 & 0.0497 & 0.0125 & 0.0157 & 0.0211 & 59.81\\
		& LightGCN 3-Layer & 0.0176 & 0.0280 & 0.0484 & 0.0122 & 0.0155 & 0.0207 & 48.54\\
		& LightGCN 2-Layer & 0.0177 & 0.0280 & 0.0482 & 0.0121 & 0.0154 & 0.0206 & 41.46 \\
		& LightGCN 1-Layer & 0.0167 & 0.0267 & 0.0460 & 0.0118 & 0.0149 & 0.0198 & 26.02 \\
		\bottomrule
	\end{tabular}
\end{table}

\subsection{Understanding the Learning of SelfCF}
In this section, we attempt to answer ``why do SelfCF framework work well for recommendation?''
Based on the line of work~\cite{bardes2022vicreg, zhang2022does}, we hypothesize that the ``stop-gradient'' design in SelfCF has the de-correlation effect on learning informative representations.
Following~\cite{bardes2022vicreg}, we define the covariance matrix of $\textbf{E} = [\textbf{e}_1, \dots, \textbf{e}_n]$ as:

\begin{equation} \label{eq:covariance}
	C(\textbf{E}) = \frac{1}{n - 1} \sum_{i=1}^{n} (\textbf{e}_{i} - \bar{\textbf{e}})(\textbf{e}_{i} - \bar{\textbf{e}})^{T}, \ \ \ \textrm{where} \ \ \ \bar{\textbf{e}} = \frac{1}{n} \sum_{i=1}^{n} \textbf{e}_{i}.
\end{equation}

The covariance regularization term $c$ is defined as the sum of the squared off-diagonal coefficients of $C(\textbf{E})$:
\begin{equation} \label{eq:cov_loss}
	c(\textbf{E}) = \frac{1}{d} \sum_{i \ne j} [C(\textbf{E})]_{i,j}^2.
\end{equation}
where $1/d$ is a scale factor. A lower covariance value indicates a better de-correlation effect on representations.

\begin{figure}[bpt]
	\centering
	\includegraphics[width=0.86\columnwidth,  trim={10 0 10 0},clip]{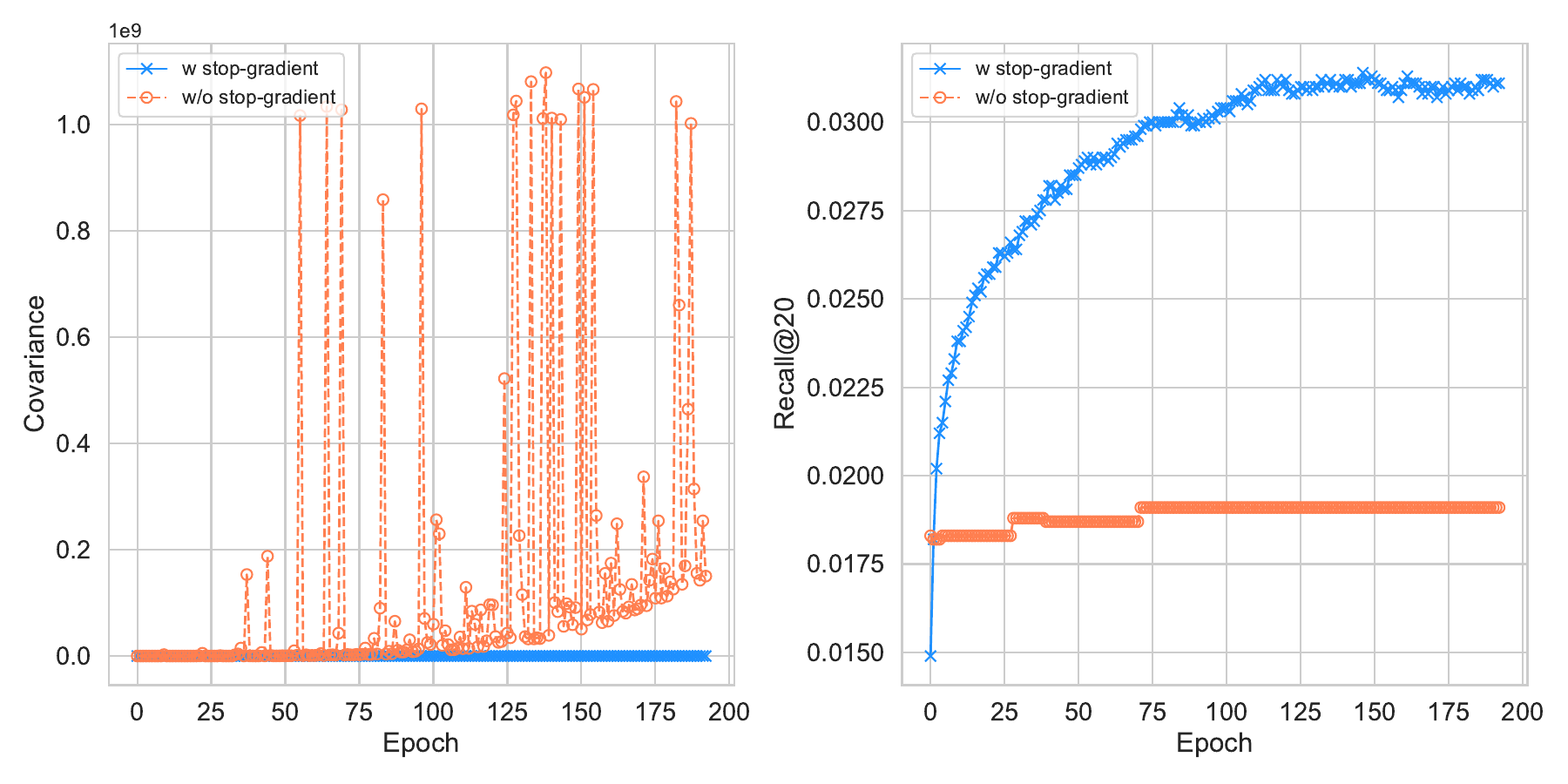} 	
	\vspace{-0.5cm}
	\caption{Comparison of SelfCF (SelfCF$_{\textrm{ed}}$) with and without ``stop-gradient'' component on Food dataset.}
	\vspace{-0.3cm}
	\label{fig:cov}
\end{figure}

Then, we compare the performance of SelfCF with and without ``stop-gradient'' component in training on Food dataset, as shown in Fig.~\ref{fig:cov}.
From the figure, we observe SelfCF with ``stop-gradient'' can decrease the covariance value of learnt representations to a significant extent compared with that of no ``stop-gradient''.
Thanks to the de-correlation effect of SelfCF, the performance on recommendation with regard to Recall@20 is consistently improved with training. 
In contrast, its performance collapses at a fixed level with the removal of the ``stop-gradient'' component.

\subsection{Hyper-parameter Sensitivity}
To guide the selection of parameters of our framework, we preform a hyper-parameter study on the performance of SelfCF.
In the implementation, we use the Food dataset as the evaluated dataset and LightGCN as the backbone of SelfCF. The results on Games and other datasets show similar patterns with Food, we put the results on Games in the Appendix for reference.
We investigate the performance changes of our framework with regard to hyper-parameters on the momentum in historical embedding, the number of layers, the ratio of embedding dropout and the proportion of edges pruned in SelfCF.

\begin{figure*}[bpt]
	\centering
	\includegraphics[width=0.9\columnwidth,  trim={10 0 10 0},clip]{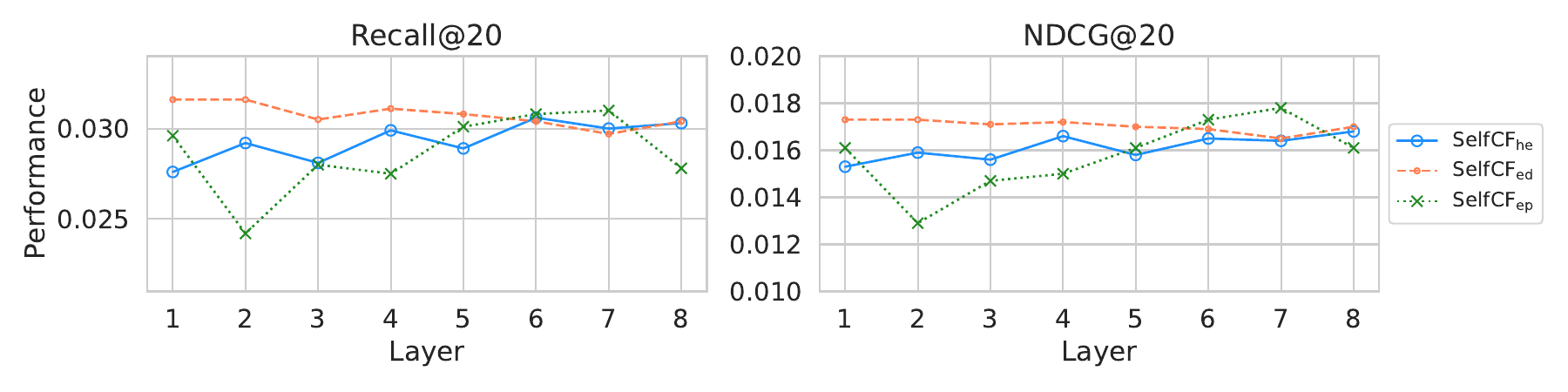}
	\vspace{-0.5cm}
	\caption{Performance of SelfCF varies with regard to the number of layers.}
	\label{fig:hypar_layer_food}
	\vspace{-0.3cm}
\end{figure*}

\textbf{The number of layers.} 
We study how layers in LightGCN affect the performance of SelfCF by stepping its range from [1, 2, 3, 4, 5, 6, 7, 8]. We plot the results in Fig.~\ref{fig:hypar_layer_food}.

SelfCF$_{\textrm{he}}$ and SelfCF$_{\textrm{ed}}$ show relatively slow performance degradation as the number of layers increasing.
The performance of SelfCF$_{\textrm{ep}}$ is not stable with regard to the number of layers.
On the contrary, the performance of SelfCF$_{\textrm{ed}}$ is not very affected with the number of layer in LightGCN.
SelfCF$_{\textrm{ed}}$ is capable of boosting up the performance of recommendation for the graph-based models within few layers.

\begin{figure}[bpt]
	\centering
	\includegraphics[width=0.99\textwidth, trim={12 0 10 0},clip]{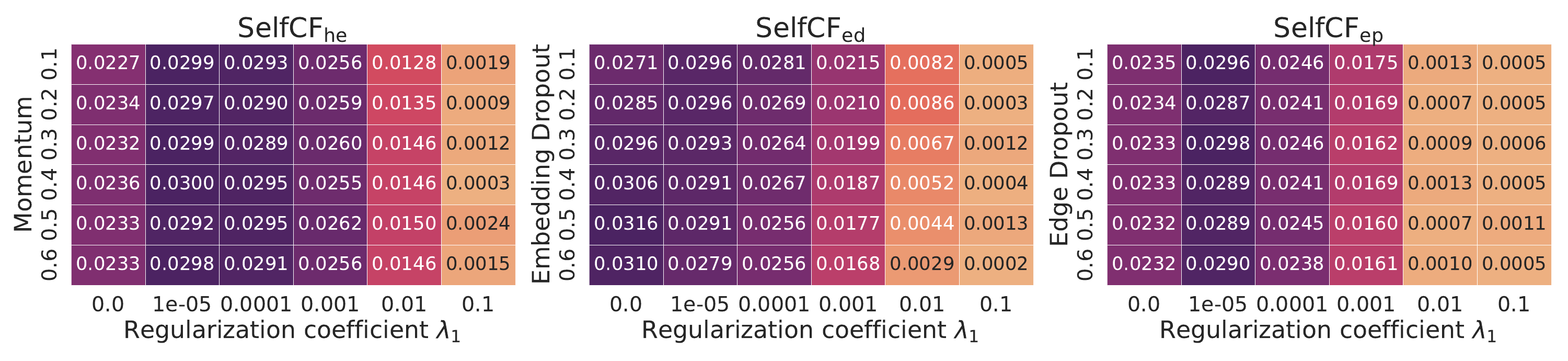}
	\vspace{-0.2cm}	
	\caption{Performance of Recall@20 for three variations of SelfCF with respect to hyper-parameters of momentum, embedding dropout, edge dropout and regularization coefficient.}
	\label{fig:hm-food-recall20}
	\vspace{-0.5cm}	
\end{figure}

\begin{figure}[bpt]
	\centering
	\includegraphics[width=0.99\textwidth, trim={12 0 10 0},clip]{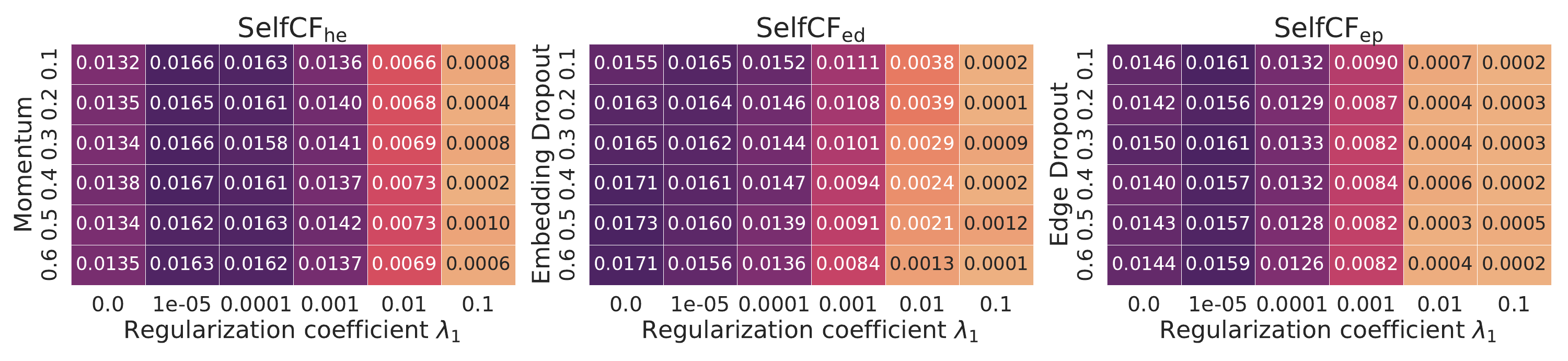}
	\vspace{-0.2cm}	
	\caption{Performance of NDCG@20 for three variations of SelfCF with respect to hyper-parameters of momentum, embedding dropout, edge dropout and regularization coefficient.}
	\label{fig:hm-food-ndcg20}
	\vspace{-0.5cm}
\end{figure}

\textbf{The momentum/dropout and regularization coefficient.} 
We set both the momentum of SelfCF$_{\textrm{he}}$ and the dropout of SelfCF$_{\textrm{ed}}$ value in the range of [0.1, 0.6] with a step of 0.1.
The $L2$ regularization coefficient $\lambda_1$ is searched in the range of \{0.0, 1e-05, 1e-04, ..., 1e-01\}.
We plot the heatmap for SelfCF$_{\textrm{he}}$, SelfCF$_{\textrm{ed}}$, SelfCF$_{\textrm{ep}}$ over Recall@20 and NDCG@20 in Fig.~\ref{fig:hm-food-recall20} and Fig.~\ref{fig:hm-food-ndcg20}, respectively.

From Fig.~\ref{fig:hm-food-recall20} and Fig.~\ref{fig:hm-food-ndcg20}, 
we observe the performance of SelfCF on Recall@20 is consistence with NDCG@20. Higher on Recall usually results in higher NDCG.
The performance of our framework is less sensitive to the momentum and dropout than the regularization factor. 
In practice, it is preferable to put weak regularization to normalize the learned embeddings.

The hyper-parameter studies also show that three variants of SelfCF exhibit similar behaviors. Hence, we analyze recommendation result and perform ablation study on SelfCF$_{\textrm{ed}}$ in the following sections.

\begin{figure}[bpt]
	\centering
	\includegraphics[width=0.76\columnwidth,  trim={10 0 10 0},clip]{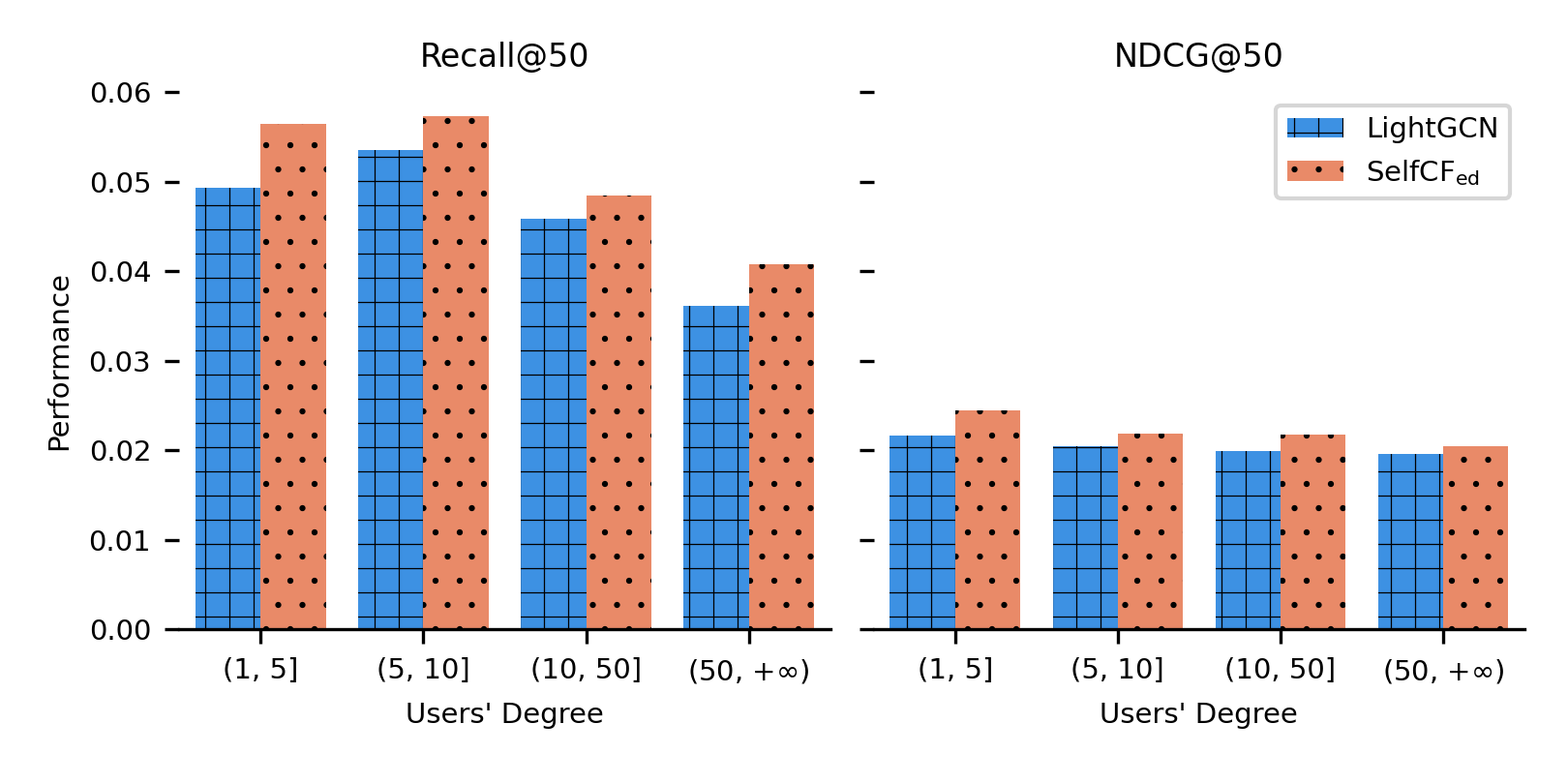} 
	\vspace{-0.5cm}
	\caption{The recommendation results on different degree of users. User' degree indicates the number of interactions of a user.}
	\label{fig:recom}
	\vspace{-0.4cm}
\end{figure}

\subsection{Diving into the Recommendation Results}
In our framework, we recommend top-$K$ items to users relying solely on positive user-item interaction pairs. 
We further study how our framework differentiate with the supervised models in recommendation results.
Specifically, we encapsulate LightGCN into our framework, and compare the recommendation results between SelfCF and LightGCN with regard to users' degree on Food.
We plot the results in Fig. \ref{fig:recom}.

On metrics Recall@50 and NDCG@50, we see SelfCF outperforms LightGCN in every category.
Our proposed framework is able to alleviate the cold-start issue to certain extent.
The most significant improvement (14.4\%) is observed on cold-start users, occupied about 63.92\% users in the testing dataset.
The second highest improvement is observed with loyal users, which gains 12.80\%.
From our data analysis, we find users with a high degree of interactions in the training are prefer to select items with a low degree in the testing.
Thus, it is difficult to recommend the right items to these users.
Our self-supervised framework can partial tackle the problem of commendation degradation on loyal user~\cite{ji2022recommender}.
We speculate the underlying reason is that for these users the supervised models sample a large number of unobserved but potentially positive items for training, which makes the models unable to consider the negatively sampled items in recommendation list.

Regarding research question \textbf{RQ2}, SelfCF boosts up the recommendation performance of all users. Especially for cold-start users, it improves the recommendation accuracy of LightGCN by 14.4\% on Recall@50.

\subsection{Representation of nodes}
\begin{figure}[bpt]
	\centering
	\begin{subfigure}[b]{.45\linewidth}
		\centering
		\includegraphics[width=0.9\linewidth, trim={60 40 50 30},clip]{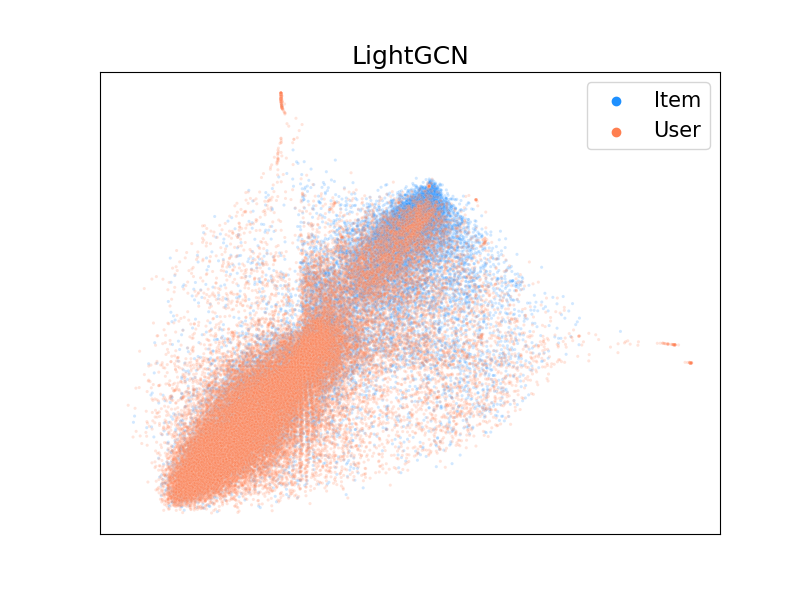}
		\label{fig:lgn_tsne}
	\end{subfigure}%
	\begin{subfigure}[b]{.45\linewidth}
		\centering
		\includegraphics[width=0.9\linewidth, trim={60 40 50 30},clip]{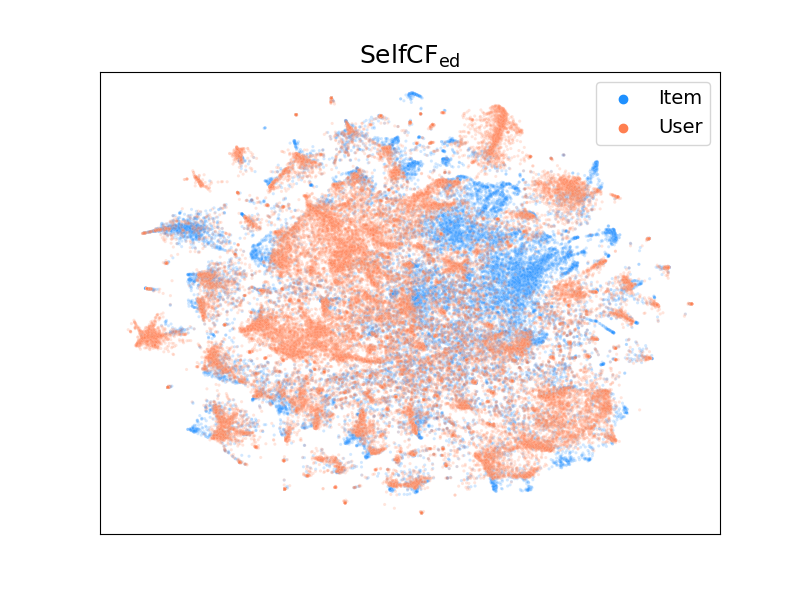}
		\label{fig:self_tsne}
	\end{subfigure}
	\caption{t-SNE plots of user and item representations learned by LightGCN and SelfCF$_{\textrm{ed}}$.}
	\label{fig:tsne}
	\vspace{-0.6cm}
\end{figure}

SelfCF leverages positive samples merely to learn the latent user and item representations. We examine how the representations differ between supervised and self-supervised learnings. 
We draw 2D t-SNE plots of node representations learned from LightGCN and SelfCF$_{\textrm{ed}}$ with regard to Food dataset in Fig~\ref{fig:tsne}.
For computational complexity consideration, we only plot the representations of users and items in the test set.
In this figure, we can observe the representations of users and items are highly melded with each other in the supervised model, LightGCN. 
On the contrary, the representations of users are items are repulsed apart in SelfCF$_{\textrm{ed}}$.
The result is in line with Equation~\ref{eq:loss_final}. 
Without the negative samples, the loss function is unable to enforce the embeddings of positive users and items similar to each other.
Instead, it maximizes the similarity subject to different conditions.
That is, our proposed self-supervised model encourages similar users congregate in a group, and similar items cluster together with each other.

\section{Ablation Study}
We investigate each component in SelfCF to study its contribution to the recommendation performance.
All ablation studies are performed on SelfCF$_{\textrm{ed}}$ trained on Food dataset.
The encapsulated baseline is LightGCN with two convolutional layers. 
The dropout of embedding for SelfCF$_{\textrm{ed}}$ is set as 0.5.

\subsection{Predictor}
We study the recommendation performance considering predictor $h$ under several variants. 
Table~\ref{tab:pred} summarizes the variants and their recommendation performance.

\begin{table}[h]
	\caption{Impact of predictor $h$.}
	\label{tab:pred}
	\begin{tabular}{lcccccc}
		\toprule
		\textbf{MLP $h$} & \textbf{R@10} & \textbf{R@20} & \textbf{R@50} & \textbf{N@10} & \textbf{N@20} & \textbf{N@50} \\
		\midrule
		1-layer MLP & 0.0198 & 0.0316 & 0.0555 & 0.0135 & 0.0173 & 0.0235 \\
		No predictor & 0.0124 & 0.0191 & 0.0359 & 0.0109 & 0.0132 & 0.0176 \\
		Fixed random init. & 0.0127 & 0.0191 & 0.0309 & 0.0109 & 0.0129 & 0.0160 \\
		2-layer MLP & 0.0199 & 0.0313 & 0.0545 & 0.0137 & 0.0173 & 0.0233 \\
		\bottomrule
	\end{tabular}
\end{table}

Different from the predictor in SimSiam~\cite{chen2021exploring}, our framework still works by removing the predictor $h$, but resulting in performance degeneration to the level of Popularity algorithm.
A fixed random initialization with the predictor makes the self-supervised framework difficult to learn good representations of users/items.
On the contrary, a 2-layer MLP also achieves a competitive performance as the 1-layer version.

\subsection{Loss function}
In contrastive learning, it is a common practice for losses measuring a cosine similarity~\cite{chent2020simple,grill2020bootstrap, wang2020understanding}. We substitute the loss function with cross-entropy similarity by modifying $\mathcal{C}$ with:
\begin{equation}
	\mathcal{C}(e_u, e_i) = -softmax(e_i) \cdot log(softmax(e_u))
\end{equation}
Table~\ref{tab:loss} shows the results compared with cosine similarity.
The cross-entropy similarity can prevent the solution collapsing to some extent. 
Cosine similarity captures the interaction preference between user and item directly, hence shows better performance.

\begin{table}
	\caption{Effectiveness of loss function.}
	\label{tab:loss}
	\begin{tabular}{lcccccc}
		\toprule
		\textbf{Similarity loss} & \textbf{R@10} & \textbf{R@20} & \textbf{R@50} & \textbf{N@10} & \textbf{N@20} & \textbf{N@50} \\
		\midrule
		Cosine & 0.0198 & 0.0316 & 0.0555 & 0.0135 & 0.0173 & 0.0235 \\
		Cross-entropy & 0.0124 & 0.0191 & 0.0350 & 0.0110 & 0.0133 & 0.0174  \\
		\bottomrule
	\end{tabular}
\end{table}

\subsection{Stop-gradient}
Existing researches~\cite{caron2020unsupervised, grill2020bootstrap, chen2021exploring, zhou2023bootstrap} on SSL highlight the crucial role of stop-gradient in preventing solution collapsing.
We evaluate with adding or removing the stop-gradient operator with/without a linear predictor. 
The results in Table~\ref{tab:sgpred} show that our self-supervised framework works even under a completely symmetry setting.
The loss function of Equation 6 is able to capture the invariant and salient features in the embeddings of users/items by dropping the noise signal.
However, without the ``stop gradient'' operator, the performance of SelfCF decreases greatly.
We speculate the loss backpropagated to both directions (online and target networks) leads to the framework unable to learn the optimal parameters of the baseline.

\begin{table}
	\caption{Effectiveness of stop-gradient (sg) operator.}
	\label{tab:sgpred}
	\begin{tabular}{lll|cccccc}
		\toprule
		\textbf{Case} & \textbf{sg} & \textbf{Predictor} & \textbf{R@10} & \textbf{R@20} & \textbf{R@50} & \textbf{N@10} & \textbf{N@20} & \textbf{N@50} \\
		\midrule
		Baseline & $\checkmark$ & $\checkmark$ & 0.0198 & 0.0316 & 0.0555 & 0.0135 & 0.0173 & 0.0235 \\
		\midrule
		(a) & - & $\checkmark$ & 0.0124 & 0.0191 & 0.0359 & 0.0109 & 0.0133 & 0.0176 \\
		(b) & - & - & 0.0124 & 0.0191 & 0.0359 & 0.0109 & 0.0132 & 0.0176 \\
		\bottomrule
	\end{tabular}
\end{table}

Based on our ablation studies with regard to research question \textbf{RQ3}, we observe that SelfCF does not rely on a single component for preventing solution collapsing. 
It shows a different behavior from other self-supervised models, in which the ``stop gradient'' operator is identified as a crucial component to prevent solution collapsing~\cite{chen2021exploring}. 
The underlying reason is that our loss function is designed as the similarity between latent embeddings of user and item, hence it can capture the preference of user to some extent.

\section{Conclusion and Future Directions}
In this paper, we propose a framework on top of Siamese networks to learn representation of users and items without negative samples or labels.
We argue the self-supervised learning techniques that widely used in vision cannot be directly applied in recommendation domain. Hence we design a Siamese network architecture that perturbs the output of backbone instead of augmenting the input data.
By encapsulating two popular recommendation models into the framework, our experiments on four datasets show the proposed framework is on par or better than other self-supervised framework, BUIR. The performance is also competitive to the supervised counterpart, obtaining a gain of 6.55\% over LightGCN.
We hope our study will shed light on further researches on self-supervised collaborative filtering.

We also discuss the potential directions based on our framework, SelfCF.
\textit{a. More powerful predictor.} In the above ablation study, we observed both 1-layer MLP and 2-layer MLP show promising performance.
Other than the supervised baselines, \eg LightGCN, the predictor is a crucial component in our framework. 
Our framework learns not only the representation of users and items but also the parameters of the predictor.
As a result, future researches can be paid to the design of predictor in our framework.
\textit{b. Combining with supervised signals.}
In recent years, a line of work integrates self-supervised learning into the classic pairwise BPR supervised loss and show promising improvement on recommendation performance~\cite{wu2021self, liu2022graph, yu2022self, yu2022graph}.
However, in our framework, we only use the positive user-item interactions pairs for recommendation. It is worth researching the integration of supervised signals into our framework.
\textit{c. Embedding augmentation methods.}
This paper proposes three embedding perturbation techniques that can be divided into two categories. 
i). Graph-based augmentation. Like BUIR, the \textbf{edge pruning} methods use another graph network to generate the contrastive embeddings.
ii). Non-graph-based augmentation. The other two techniques directly distort the original view and can significantly ease the computation burden.
However, other methods~\cite{liu2022graph} for embedding augmentation can be explored in SelfCF. \textit{d. Fusing multimodal features for effective recommendation.} To alleviate the data sparsity problem and the cold start issue in CF, various methods have been developed to fuse the multimodal information (\eg text descriptions and images) of items into the current CF paradigm~\cite{zhou2023comprehensive}. In this direction of future work, we are interested in exploring effective ways of fusing multimodal features for recommendation.

\begin{acks}
	This work is partially supported by the MOE AcRF Tier 1 funding (RG90/20), Alibaba Group through Alibaba Innovative Research (AIR) Program and Alibaba-NTU Singapore Joint Research Institute (JRI), Nanyang Technological University, Singapore.
	We thank Professor Toru Ishida from	the Department of Computer Science of Hong Kong Baptist University for his valuable advices on this work.
\end{acks}

\bibliographystyle{ACM-Reference-Format}
\bibliography{selfcf}

\appendix

\section{Hyper-parameter study on Games}
\begin{figure*}[bpt]
	\centering
	\includegraphics[width=0.95\columnwidth,  trim={10 0 10 0},clip]{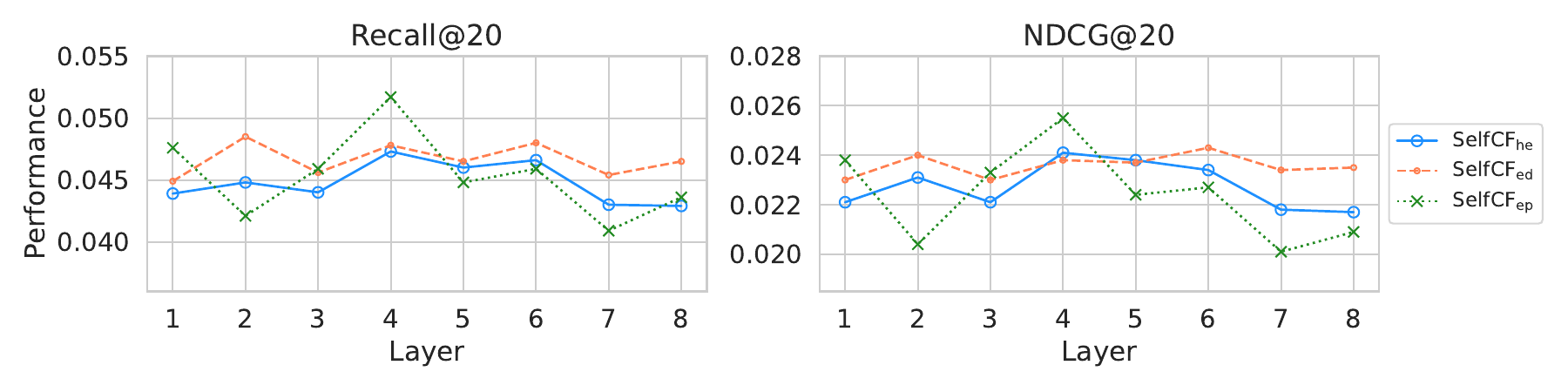}	
	\vspace{-0.4cm}
	\caption{Performance of SelfCF varies with regard to the number of layers on Games.}	
	\vspace{-0.4cm}
	\label{fig:hypar_layer_games}
\end{figure*}

\begin{figure}[bpt]
	\centering
	\includegraphics[width=0.95\textwidth, trim={12 0 10 0},clip]{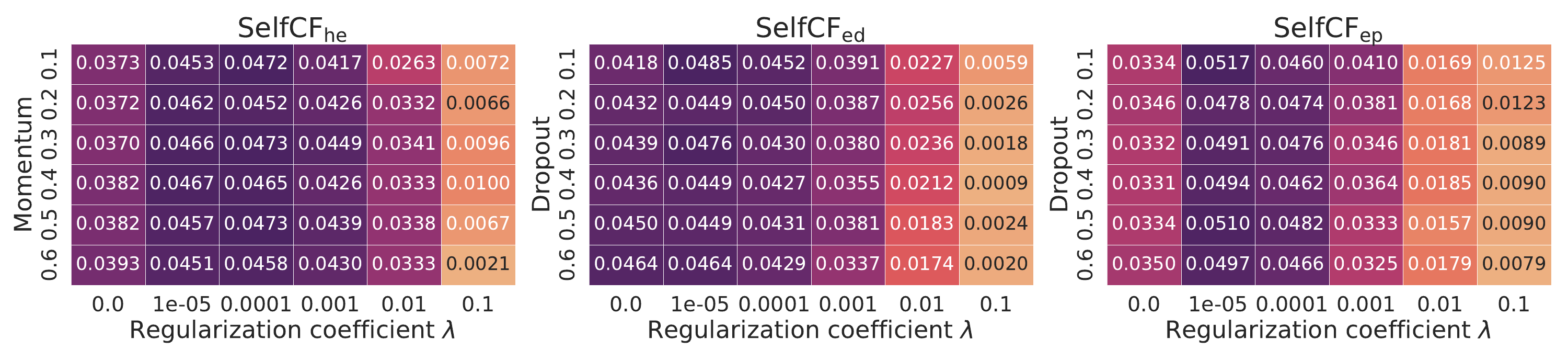}
	\vspace{-0.4cm}
	\caption{Performance of Recall@20 for three variations of SelfCF with respect to hyper-parameters of momentum, embedding dropout, edge dropout and regularization coefficient on Games.}
	\vspace{-0.4cm}
	\label{fig:hm-games-recall20}
\end{figure}

\begin{figure}[bpt]
	\centering
	\includegraphics[width=0.95\textwidth, trim={12 0 10 0},clip]{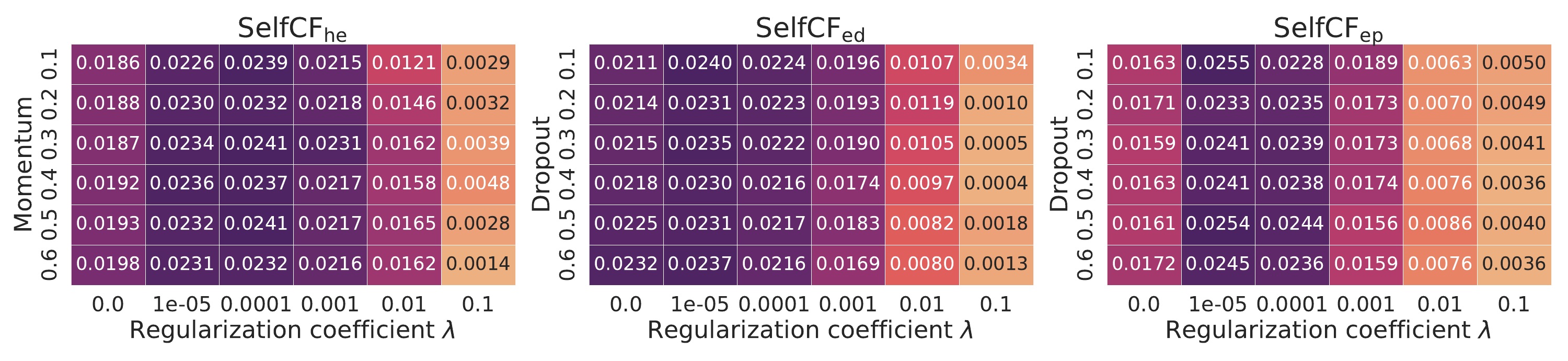}
	\vspace{-0.4cm}
	\caption{Performance of NDCG@20 for three variations of SelfCF with respect to hyper-parameters of momentum, embedding dropout, edge dropout and regularization coefficient on Games.}
	\vspace{-0.2cm}
	\label{fig:hm-games-ndcg20}
\end{figure}

We plot the performance of SelfCF varies with the number of layers in Fig.~\ref{fig:hypar_layer_games}. The performance of Recall@20 and NDCG@20 changes with momentum, embedding dropout, edge dropout and regularization coefficient in Fig.~\ref{fig:hm-games-recall20} and Fig.~\ref{fig:hm-games-ndcg20}, respectively.
The patterns in Games are in line with that of Food.

\section{Performance of LightGCN under different sampling methods}
In response to the research issue mentioned in Introduction, we evaluate the performance of LightGCN under both uniform sampling (UniS) and dynamic negative sampling (DNS) methods on MOOC.
The results are summarized in Table~\ref{tab:lightgcn-sampling}.
DNS samples hard negative user-item pairs for LightGCN, hence obtains higher ranking score (\ie NDCG@10) when $K$ is low in top-$K$.
However, when $k$ increases, it difficult to retrieve related but low ranked items for a target user. Because it always rank items based on the current representations of users and items learned by LightGCN. In many cases, the sampled negative pairs are from the test set.
As a result, the recall of dynamic negative sampling on LightGCN with regard to $K=20$ and $K=50$ is worse than the uniform sampling method.
\begin{table}
	\caption{Influence of sampling methods on LightGCN evaluated with MOOC dataset.}
	\label{tab:lightgcn-sampling}
	\begin{tabular}{lcccccc}
		\toprule
		\textbf{Model} & \textbf{R@10} & \textbf{R@20} & \textbf{R@50} & \textbf{N@10} & \textbf{N@20} & \textbf{N@50} \\
		\midrule
		LightGCN-UniS & 0.2507 & 0.3321 & 0.4844 & 0.1588 & 0.1835 & 0.2208\\
		LightGCN-DNS & 0.2560 & 0.3297 & 0.4644 & 0.1644 & 0.1864 & 0.2191\\
		\midrule
		SelfCF$_{\textrm{he}}$ & 0.2545 & 0.3500 & 0.4914 & 0.1696 & 0.1986 & 0.2328 \\
		SelfCF$_{\textrm{ed}}$ & 0.2460 & 0.3337 & 0.5088 & 0.1752 & 0.2009 & 0.2443 \\
		SelfCF$_{\textrm{ep}}$ & 0.2514 & 0.3485 & 0.4964 & 0.1671 & 0.1963 & 0.2323 \\	
		\bottomrule
	\end{tabular}
	\vspace{-15pt}
\end{table}

\end{document}